\documentstyle[onecolumn]{mn}

\newcounter{parentequation}

\def\pmb#1{\setbox0=\hbox{#1}%
    \kern-.025em\copy0\kern-\wd0
    \kern.05em\copy0\kern-\wd0
    \kern-.025em\raise.0433em\box0}

\def\ltsima{$\; \buildrel < \over \sim \;$}
\def\gtsima{$\; \buildrel > \over \sim \;$}
\def\simlt{\lower.5ex\hbox{\ltsima}}
\def\simgt{\lower.5ex\hbox{\gtsima}}
\def\p2Y{\;_2Y}
\def\m2Y{\;_{-2}Y}

\def\etal{{\it et al.}\rm}
\def\etals{{\it et al. }\rm}
\def\mk2{\mu {\rm K}^2}
\def\planck{\it Planck\rm}
\def\plancks{\it Planck \rm}

\begin{document}

\title[B-mode detection at low multipoles]
{Impact of Galactic polarized emission on B-mode detection
at low multipoles}

\author[Efstathiou, Gratton and Paci]{G. Efstathiou$^{1}$,  S. Gratton$^{1}$
and F. Paci$^{1,2,3,4}$\\
1. Kavli Institute for Cosmology Cambridge and 
Institute of Astronomy, Madingley Road, Cambridge, CB3 OHA.\\
2. Dipartimento di Astronomia, Universit\`a degli Studi di Bologna, via
Ranzani 1, I-40127 Bologna, Italy. \\
3. INAF/IASF-BO, Istituto di Astrofisica Spaziale e Fisica Cosmica di Bologna
via Gobetti 101, I-40129 Bologna, Italy.\\
4.INFN, Sezione di Bologna, Via Irnerio 46, I-40126 Bologna, Italy. }

\maketitle

\begin{abstract}
  We use a model of polarized Galactic emission developed by the the
  Planck collaboration to assess the impact of foregrounds on B-mode
  detection at low multipoles. Our main interest is to applications of
  noisy polarization data and in particular to assessing the
  feasibility of $B$-mode detection by \planck. This limits the
  complexity of foreground subtraction techniques that can be applied
  to the data.  We analyze internal linear combination techniques and
  show that the offset caused by the dominant $E$-mode polarization
  pattern leads to a fundamental limit of $r \sim 0.1$ for the
  tensor-scalar ratio even in the absence of instrumental noise. We
  devise a simple, robust, template fitting technique using
  multi-frequency polarization maps.  We show that template fitting
  using \plancks data alone offers a feasible way of recovering
  primordial $B$-modes from dominant foreground contamination, even in
  the presence of noise on the data and templates. We implement and
  test a pixel-based scheme for computing the likelihood function of
  cosmological parameters at low multipoles that incorporates
  foreground subtraction of noisy data.

\vskip 0.1 truein


\vskip 0.3 truein

\end{abstract}

\section{Introduction}

In the last decade, observations of cosmic microwave background (CMB)
anisotropies have provided one of the most powerful probes of
cosmology.  By combining CMB anisotropy data with a variety of other
data, many of the key parameters that define our Universe have been
determined with unprecedented precision (see {\it e.g.} Komatsu \etal,
2008, and references therein). Nevertheless, many important questions
remained unanswered. One of the most important is the amplitude of a
`B-mode' polarization signature in the CMB. Scalar perturbations
generate during inflation generate purely a divergence-like $E$-mode
polarization pattern in the CMB, whereas tensor perturbations would
produce a distinctive curl-like ($B$-mode) polarization signature
together with an $E$-mode pattern of roughly equal
amplitude\footnote{Gravitational lensing of CMB $E$-modes by
  intervening matter generates a $B$-mode anisotropy (see Lewis and
  Challinor, 2006, for a review). This effect will be ignored in this
  paper, since we will concentrate on the detectability of tensor
  modes at low multipoles $\ell \simlt 20$, where the effects of
  lensing are small.}(Zaldarriaga and Seljak, 1997; Kamionkowski,
Kosowsky and Stebbins, 1997).

A detection of a primordial $B$-mode anisotropy would provide crucial
evidence that inflation actually took place. Furthermore, a
measurement of the relative amplitude of the tensor and scalar
primordial power spectra (the tensor-scalar ratio $r$, see Peiris
\etal, 2003, for a precise definition) would fix the energy scale of
inflation via
\begin{equation}
 V^{1/4} \approx 3.3 \times 10^{16} r^{1/4} \; {\rm GeV}, \label{I1}
\end{equation}
(Lyth 1984) providing critical constraints on inflationary models (see
Baumann \etals 2008, for a review). It is therefore no surprise that a
number of sensitive ground-based/sub-orbital CMB polarization
experiments are either planned or in progress. Examples include {\it
  BICEP} (Yoon \etal, 2006), {\it CLOVER} (North \etal, 2008), {\it
  EBEX} (Oxley \etal, 2004),  {\it QUIET} (Seiffert \etal, 2006) and {\it
  SPIDER} (Crill \etal, 2008). In addition, groups in  Europe and
the US have considered designs for a $B$-mode optimised space satellite
capable of probing tensor-scalar ratios $r \simlt 10^{-2}$ (de
Bernardis \etals 2008; Bok \etals 2008).

The {\it Planck} satellite\footnote{For a description of \plancks
 and its science case, see {\it `The Scientific
    Programme of Planck'} (2005), hereafter SPP05.} is scheduled for
launch in April 2009 and has polarization sensitivity in $7$ channels
over the frequency range $30$-$353$ GHz. As described in SPP05, the
sensitivity of \plancks limits $B$-mode recovery to low multipoles
$\ell \simlt 20$\footnote{Although in theory it may be possible to extract
some information on primordial $B$-modes from the noise dominated data at
higher multipoles, a number of systematic effects such as cross-polar
leakage and errors in the polarizer angles are expected to
dominate at higher multipoles.}. Nevertheless, until a new polarization-optimised
satellite is flown, \plancks is the only experiment capable of probing
these low multipoles. It is therefore important to analyse \planck's
performance for $B$-mode detection in the presence of realistic
noise levels and polarized foregrounds and therefore to assess whether
it can provide useful complementary data to experiments probing higher
multipoles. This is the main goal of this paper.

The problem of detecting primordial $B$-modes at low multipoles is
unusually difficult. Unlike the detection of temperature anisotropies,
Galactic foregrounds are expected to have a much larger amplitude than
any putative primordial $B$-mode signal over the entire sky (see
Section 2). Accurate foreground removal is therefore essential for
$B$-mode detection at low multipoles. There has been an enormous
amount of work on CMB foreground subtraction (see {\it e.g.} the
review by Delabroille and Cardoso 2007; Leach \etals 2008). A variety
of methods have been developed, ranging from `blind' techniques that
make few physical assumptions concerning the Galactic foregrounds
(examples include Internal Linear Combination (ILC), {\it e.g.}
Bennett \etals (2003), Independent Component Analysis {\it e.g.}
Hyv\"arinen (1999) and its fast and spectral-matching variants (Maino
\etals 2002; Delabrouille, Cardoso and Patanchon 2003)), `semi-blind'
methods that make limited use of prior information on the foregrounds
(such as Maximum Entropy, Stolyarov \etals (2002) and template
matching (Bennett \etals 2003; Slosar, Seljak and Makarov 2004; Slosar
and Seljak 2004; Eriksen \etals 2004a)) and parametric fitting
techniques based on physical models of the foregrounds (Eriksen \etals
2006; Eriksen \etals 2008; Dunkley \etals 2008a).  A number of other
methods have been developed which incorporate some aspects of these
techniques and use, for example, wavelet or harmonic decompositions
({\it e.g.}  Tegmark, de Oliviera-Costa and Hamilton 2003; Hansen
\etals 2006; Norgaard-Nielsen and Jorgensen 2008). Some of these
methods provide approximations to the likelihood function for
cosmological parameters ({\it eg.} Eriksen \etals 2008) and there has
also been some work (Gratton 2008) directly addressing the question of
modelling the likelihood function from multi-frequency maps.  Almost
all of these methods have been developed for temperature foreground
subtraction. In contrast to the temperature data, the noise level of
\plancks polarization maps will be high. There is therefore limited
information in polarization and hence a restriction on the complexity
of polarized foreground removal algorithms that the data can support.

The problem of $B$-mode detection in the presence of Galactic
foregrounds has been considered by a number of authors. Tucci \etals
(2005) performed a Fisher matrix analysis for idealised experiments
including foregrounds. Amblard, Cooray and Kaplinghat (2007)
investigated harmonic ILC subtraction for $B$-mode detection at high
mulipoles ($\ell \simgt 20$) for various experimental configurations.
Betoulle \etals (2009) considered the application of Spectral Matching
Independent Component Analysis (SMICA) to perform a Fisher matrix
analysis for various experiments, including \planck. The work most
closely related to ours is the paper by Dunkley \etals (2008b), which
focuses on $B$-mode detection with a future satellite with high
signal-to-noise in polarization, rather than the low signal-to-noise
case relevant to \planck. Low signal-to-noise introduces additional
complexity to the foreground subtraction problem, nevertheless, there
are strong similarities between our approaches.

The layout of this paper is as follows.  The Planck Sky Model, which
is used in this paper to model polarized Galactic foregrounds, is
described briefly in Section 2. This model is compared to realizations
of the primordial CMB polarization signature to determine the
magnitude of the foreground subtraction problem. ILC foreground
subtraction is described in Section 3. We show that the ILC method is
fundamentally limited for $B$-mode detection because of the offset
associated with the dominant $E$-mode signal. Section 4 analyses
foreground template subtraction techniques and we present a heuristic
model for constructing a pixel-based polarization likelihood
function. This is applied to simulations with \planck-like noise.
Section 5 introduces a classification scheme for foreground
subtraction methods based on their dominant errors. Our conclusions
are summarized in Section 6.

\section{The Magnitude of the Problem}

\subsection{The Planck Sky Model}

The Planck Sky Model (PSM) has been developed by the Planck Component
Separation Working Group for use in simulations of the \plancks
mission. Summaries of the model are given by Leach \etals (2008) and
Dunkley \etals (2008b) and a detailed description will be provided in
a forthcoming paper by Delabrouille \etals (2009, in preparation). The
polarized foreground model used in this paper is similar to that used
by Dunkley \etals (2008b). Briefly, the model includes polarization
from a power-law synchrotron component with geometrical suppression
factors, polarization angles and polarization fractions based on the
magnetic field model of Miville-Deschenes \etals (2008). It also
includes a power-law dust component based on an IRAS dust template
derived by Finkbeiner, Davis and Schlegel (1999). Polarization from
point sources is ignored. The resulting dust polarization fraction in
this model is $\sim 5\%$ over most of the sky, corresponding to the
`high' polarization fraction used in Dunkley \etals (2008b).
Preliminary indications from the BICEP experiment suggest a
significantly lower polarization fraction of $\sim 1$-$2\%$ close to
the Galactic plane, though this figure may be unrepresentative of
regions at higher Galactic latitude where depolarization may be lower
(BICEP collaboration, private communication).

\begin{figure*}
\vskip 8.7 truein

\includegraphics{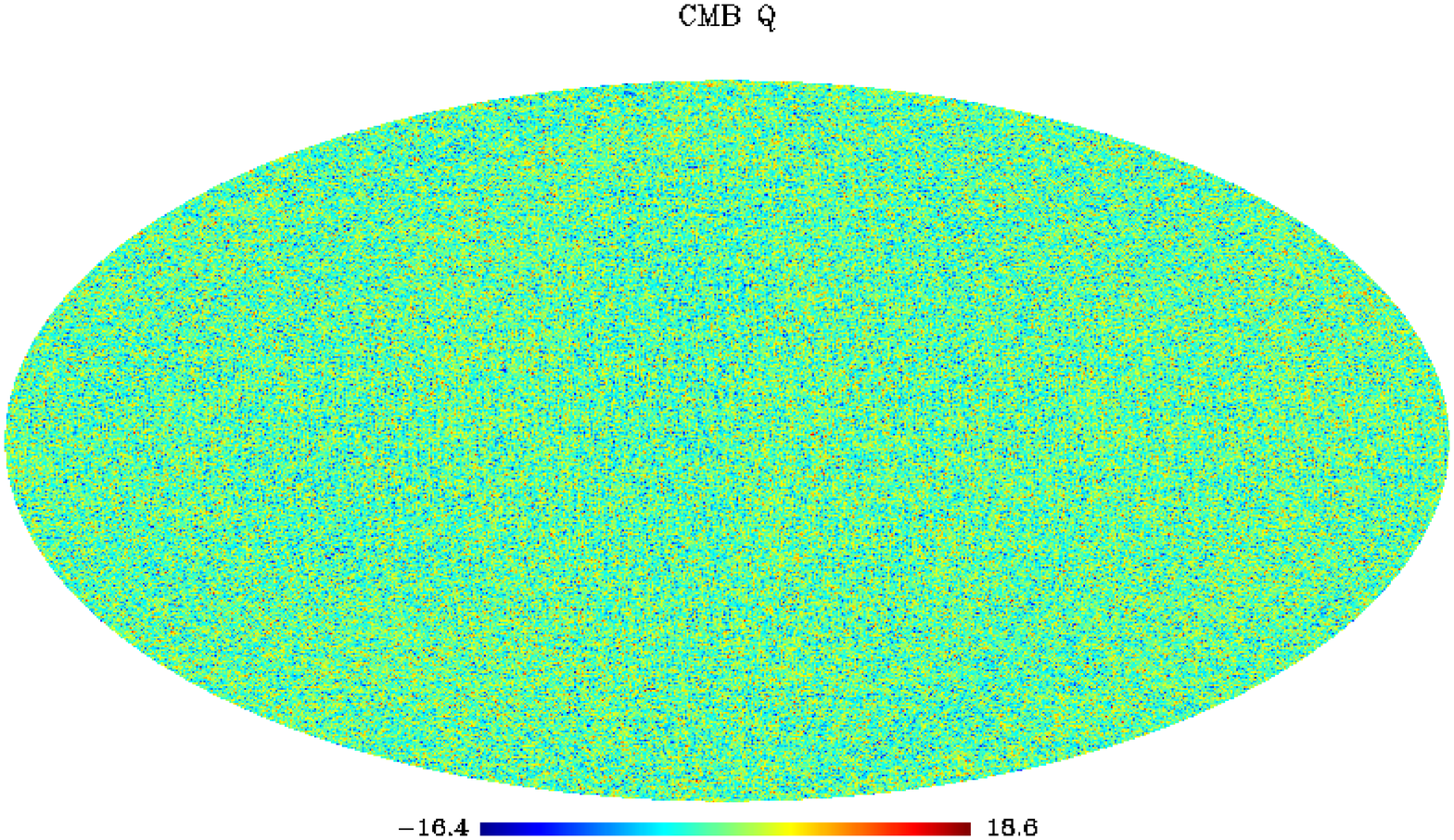}
\includegraphics{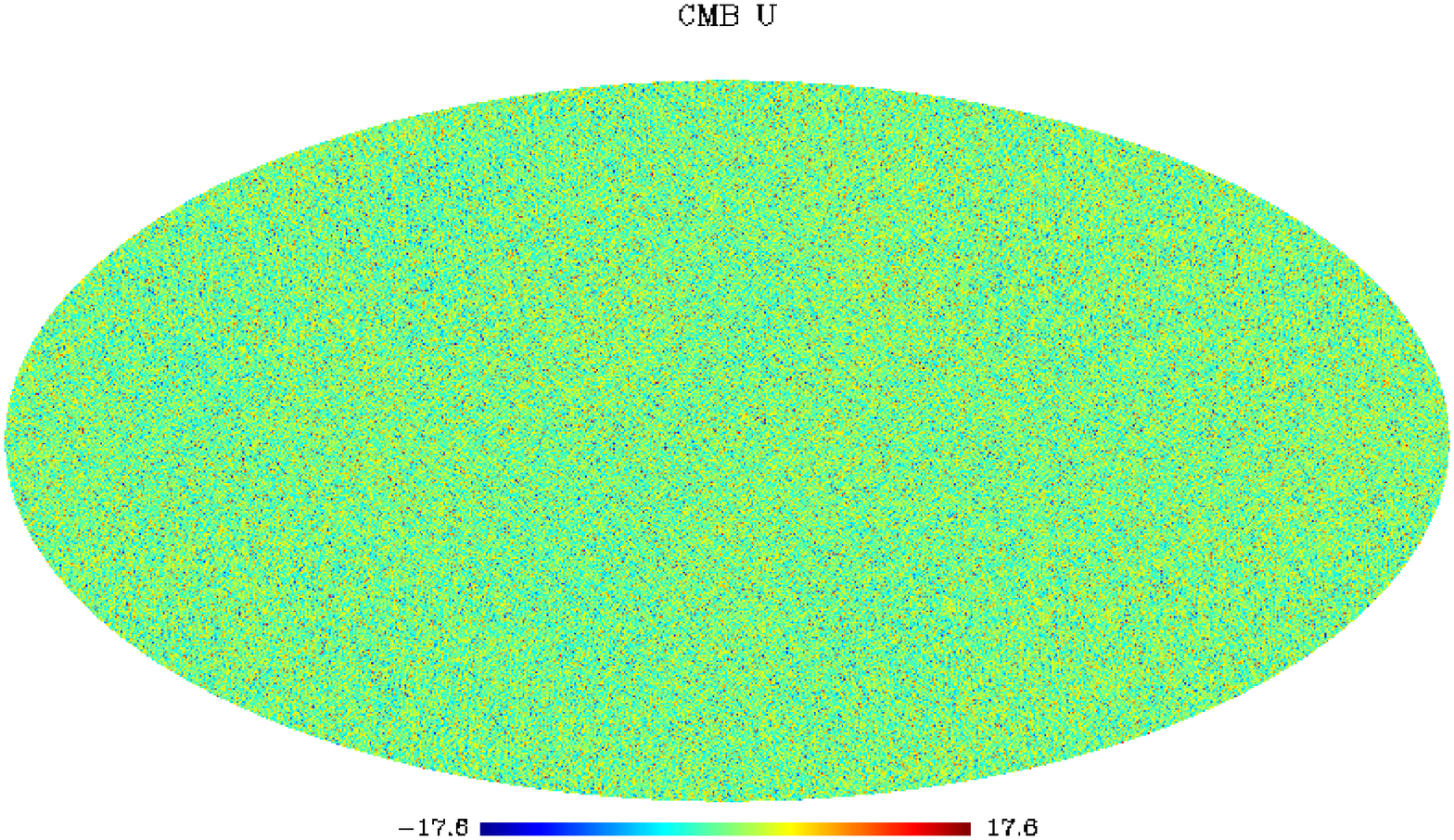}
\includegraphics{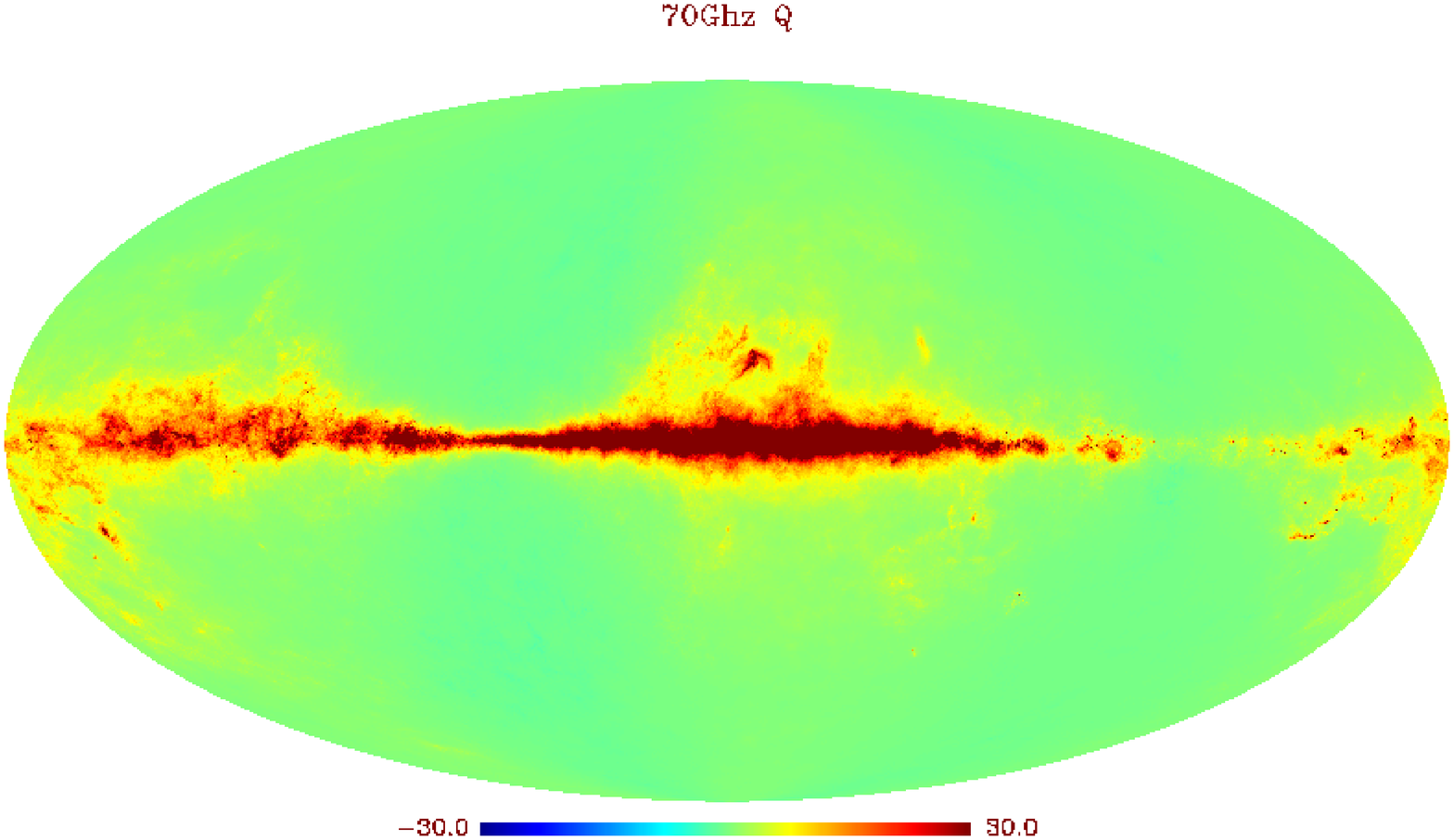}
\includegraphics{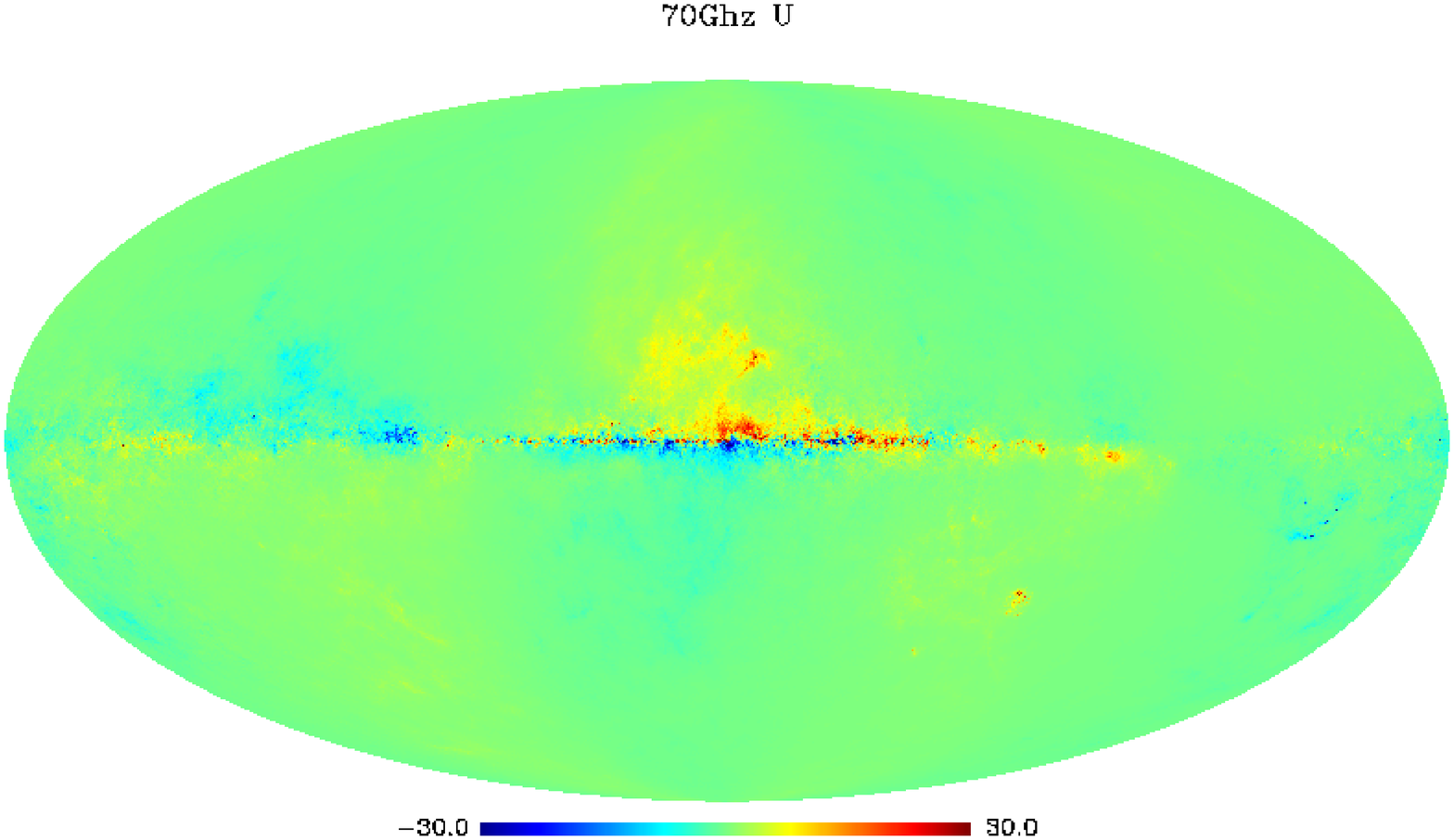}
\includegraphics{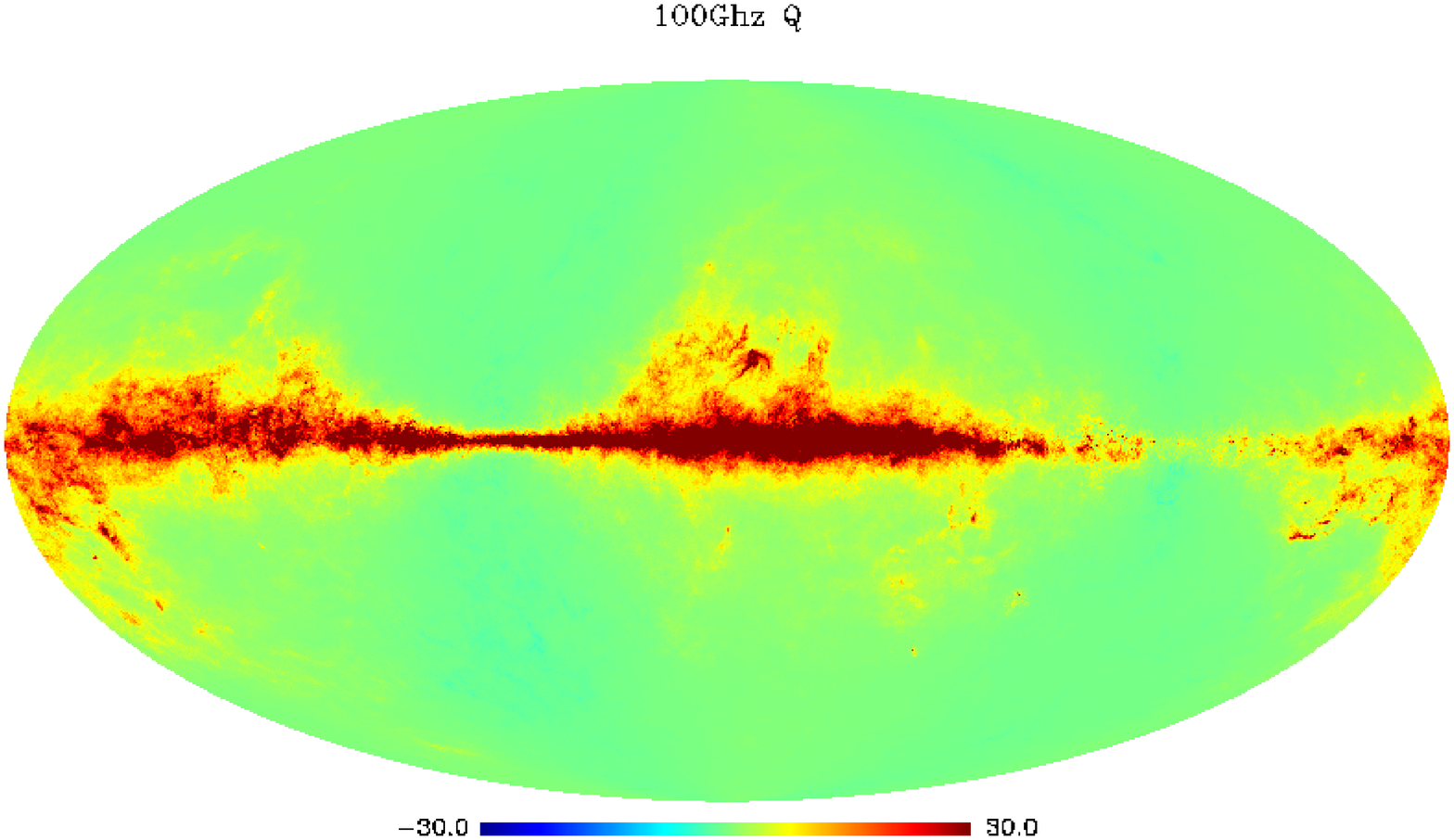}
\includegraphics{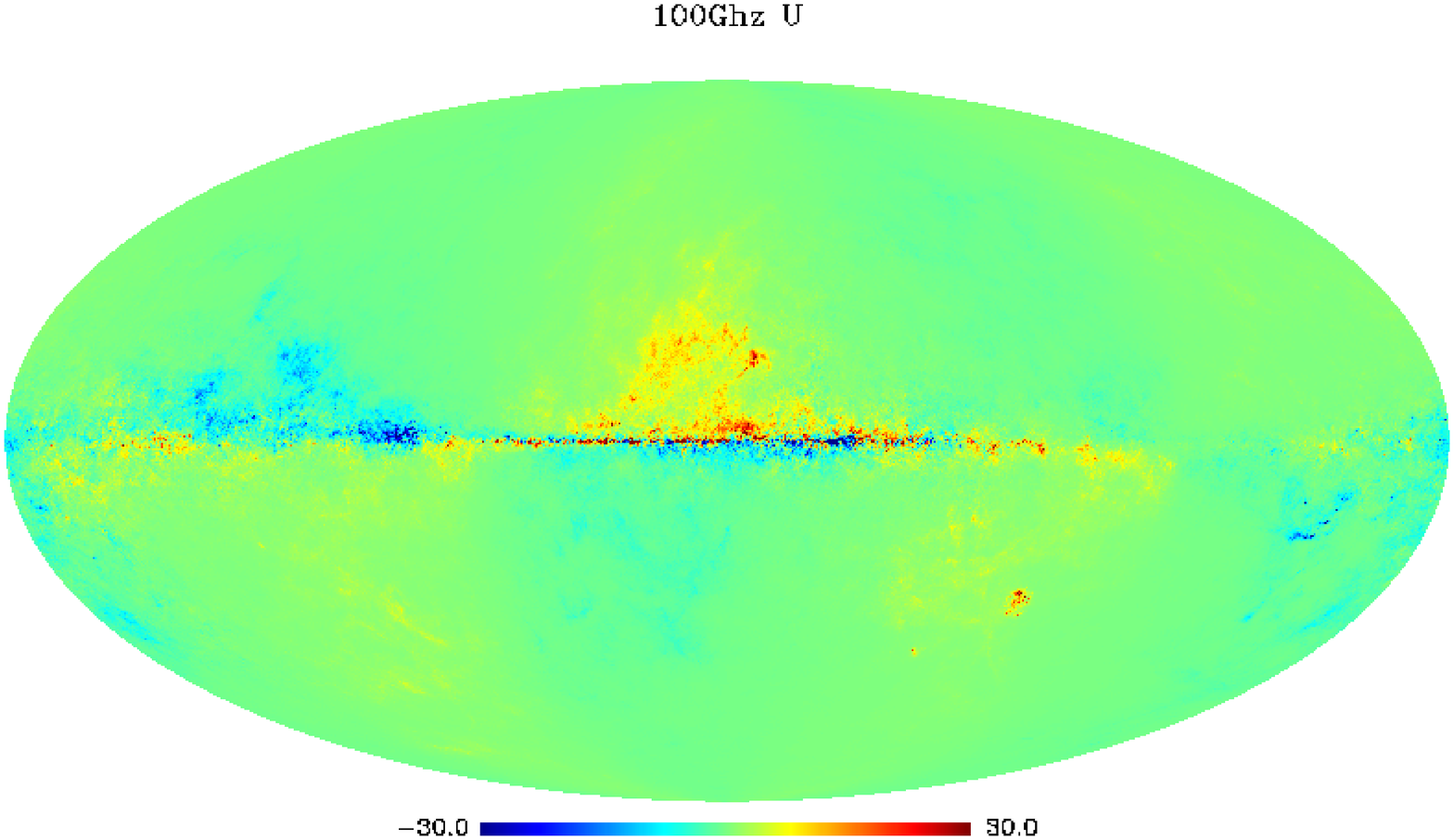}
\includegraphics{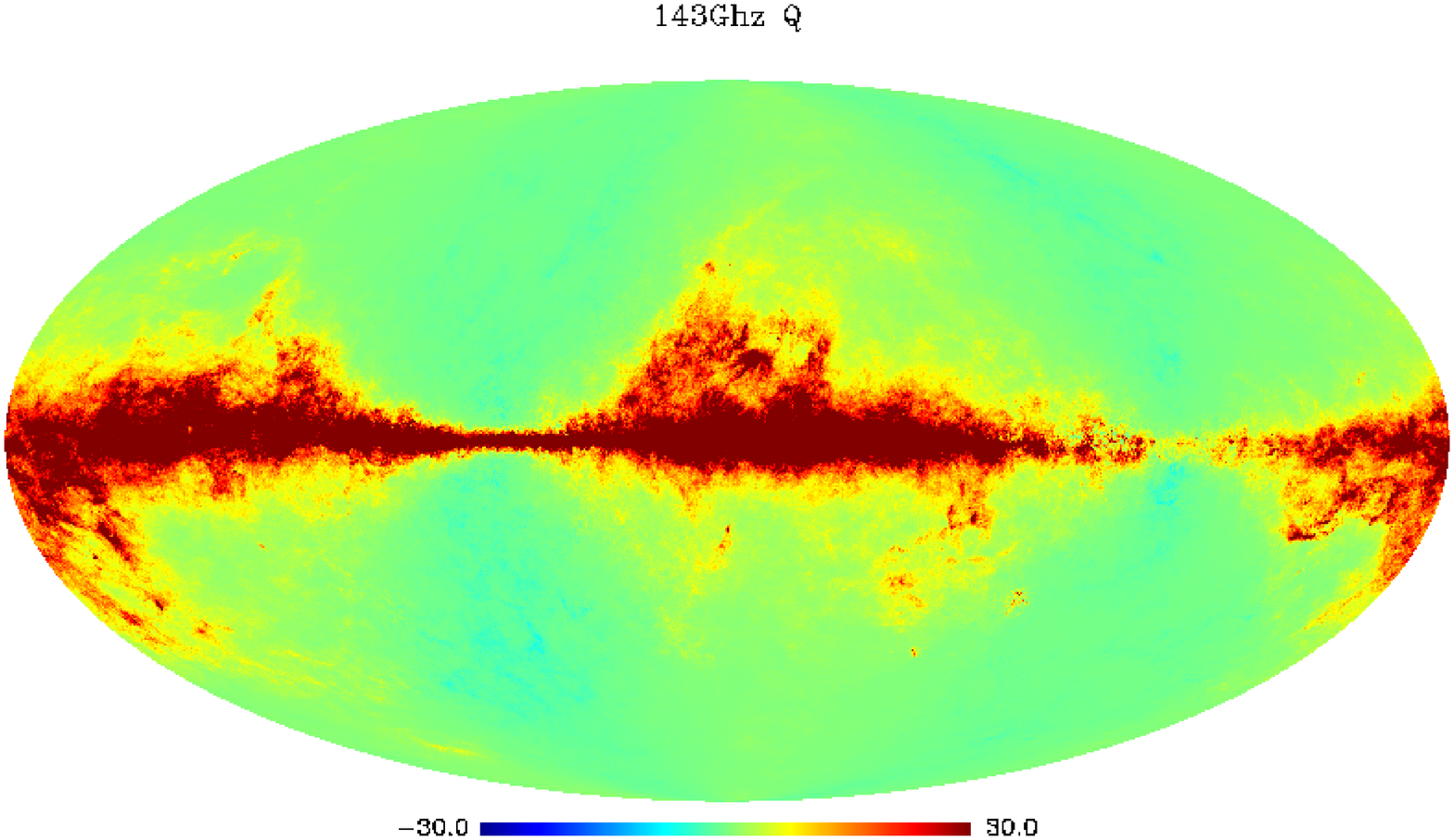}
\includegraphics{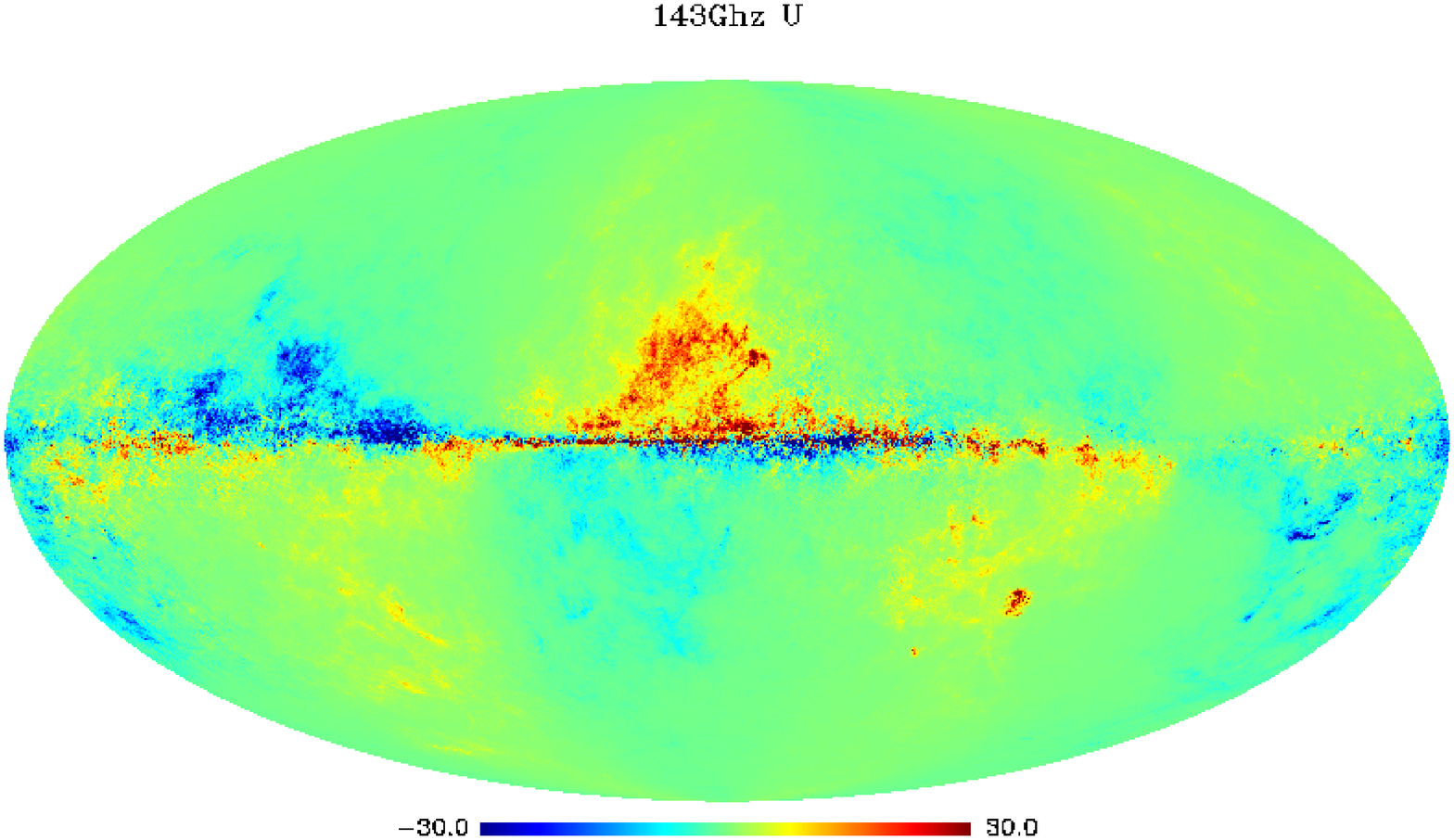}
\includegraphics{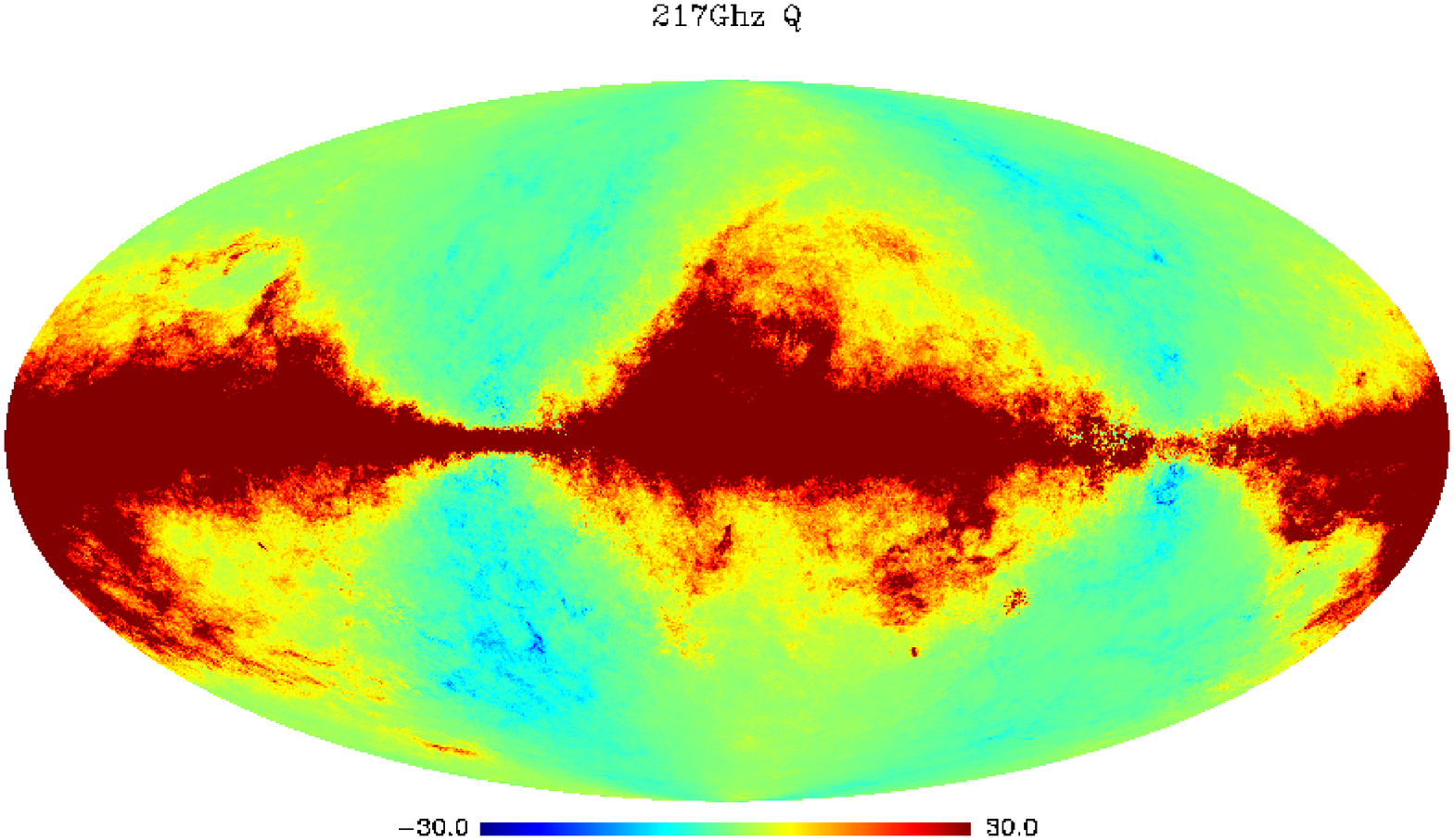}
\includegraphics{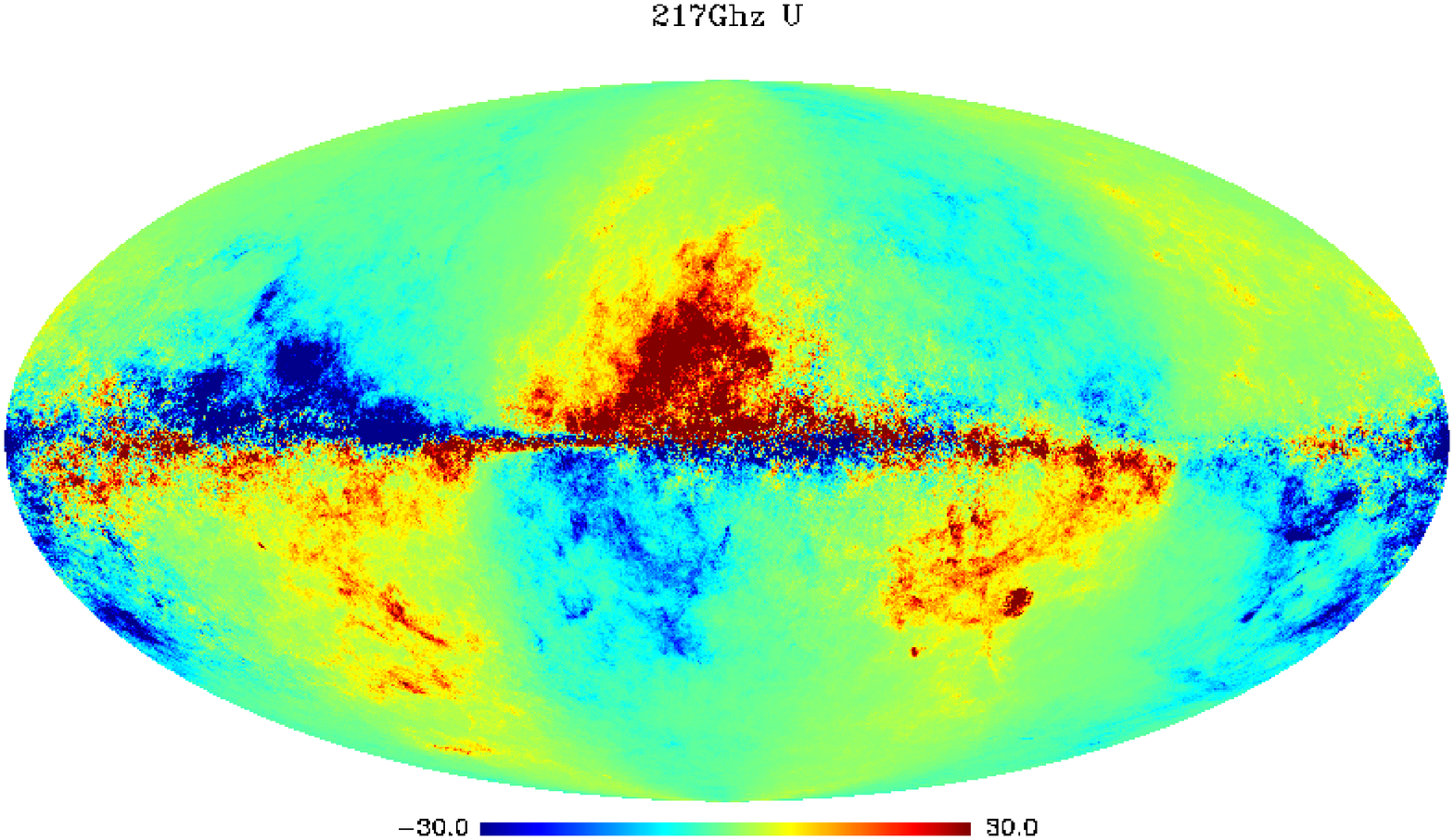}

\caption
{Q (left) and U (right) maps: the upper panel shows CMB simulations. The remaining 
panels show the PSM at $70$, $100$, $143$ and $217$ GHz. The temperature scale
(thermodynamic temperature) is in $\mu {\rm K}$. All maps were generated
at Healpix NSIDE=2048.}

\label{figure1}
\end{figure*}

\begin{figure*}
\vskip 8.7 truein

\includegraphics{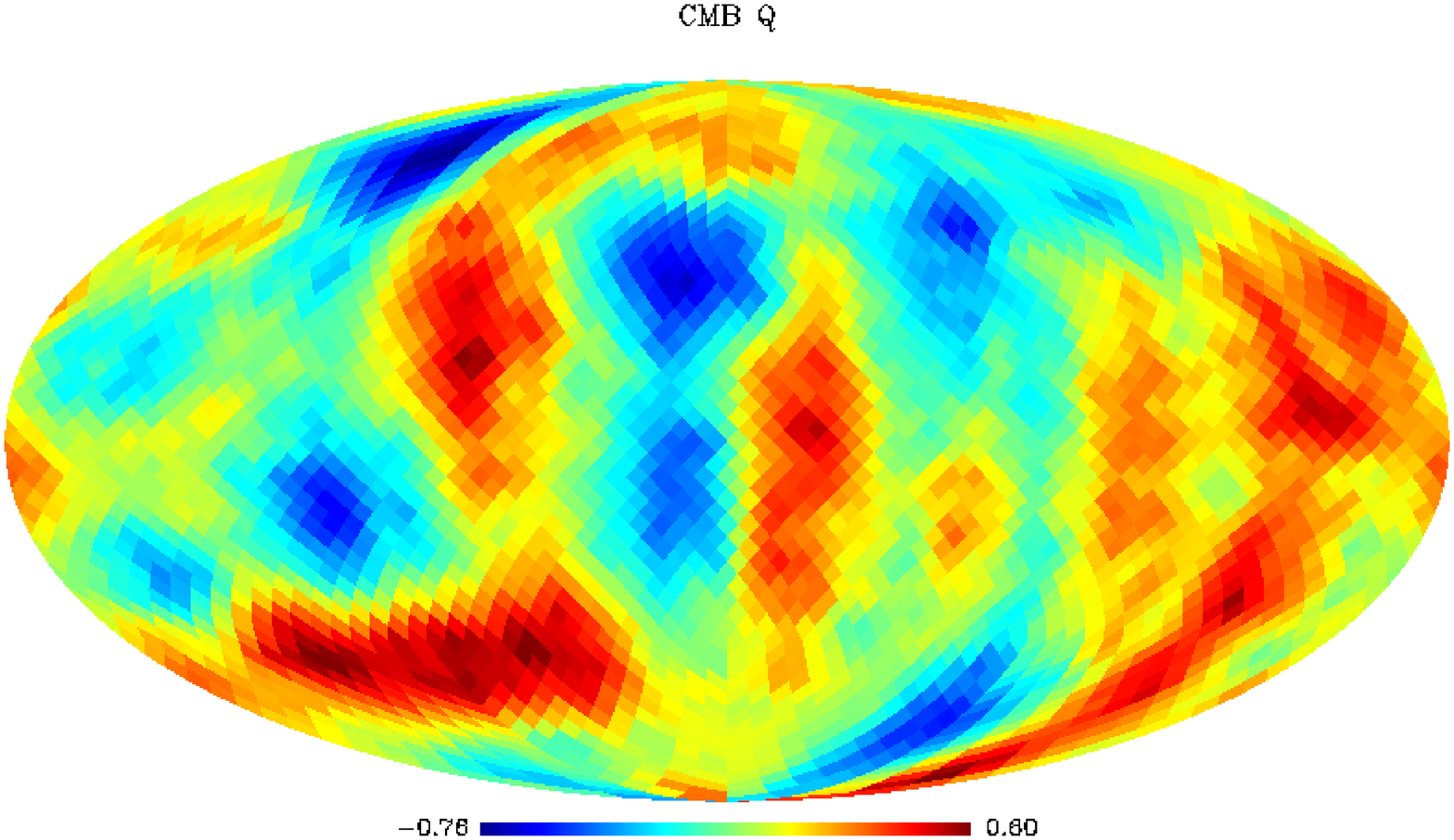}
\includegraphics{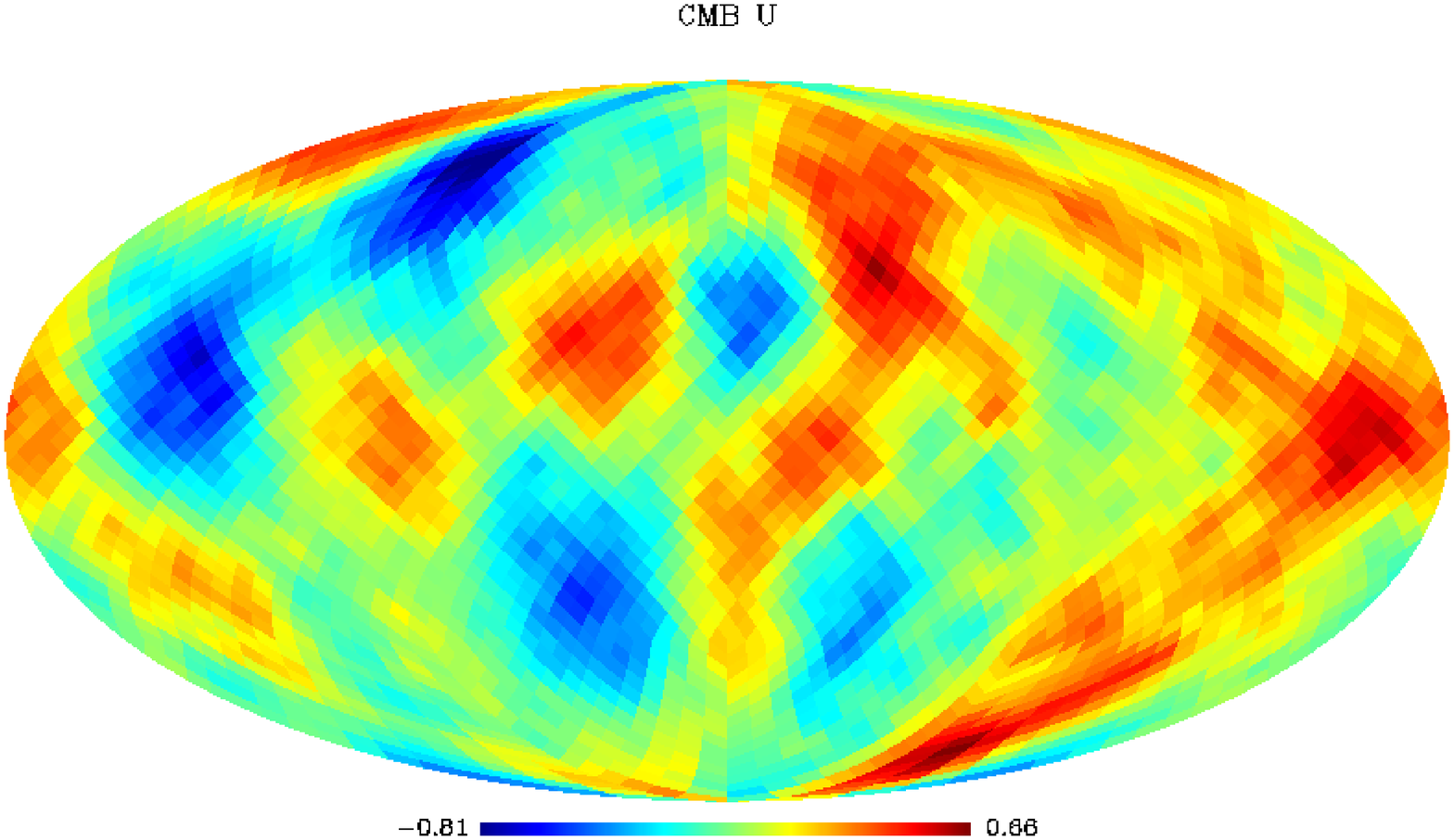}
\includegraphics{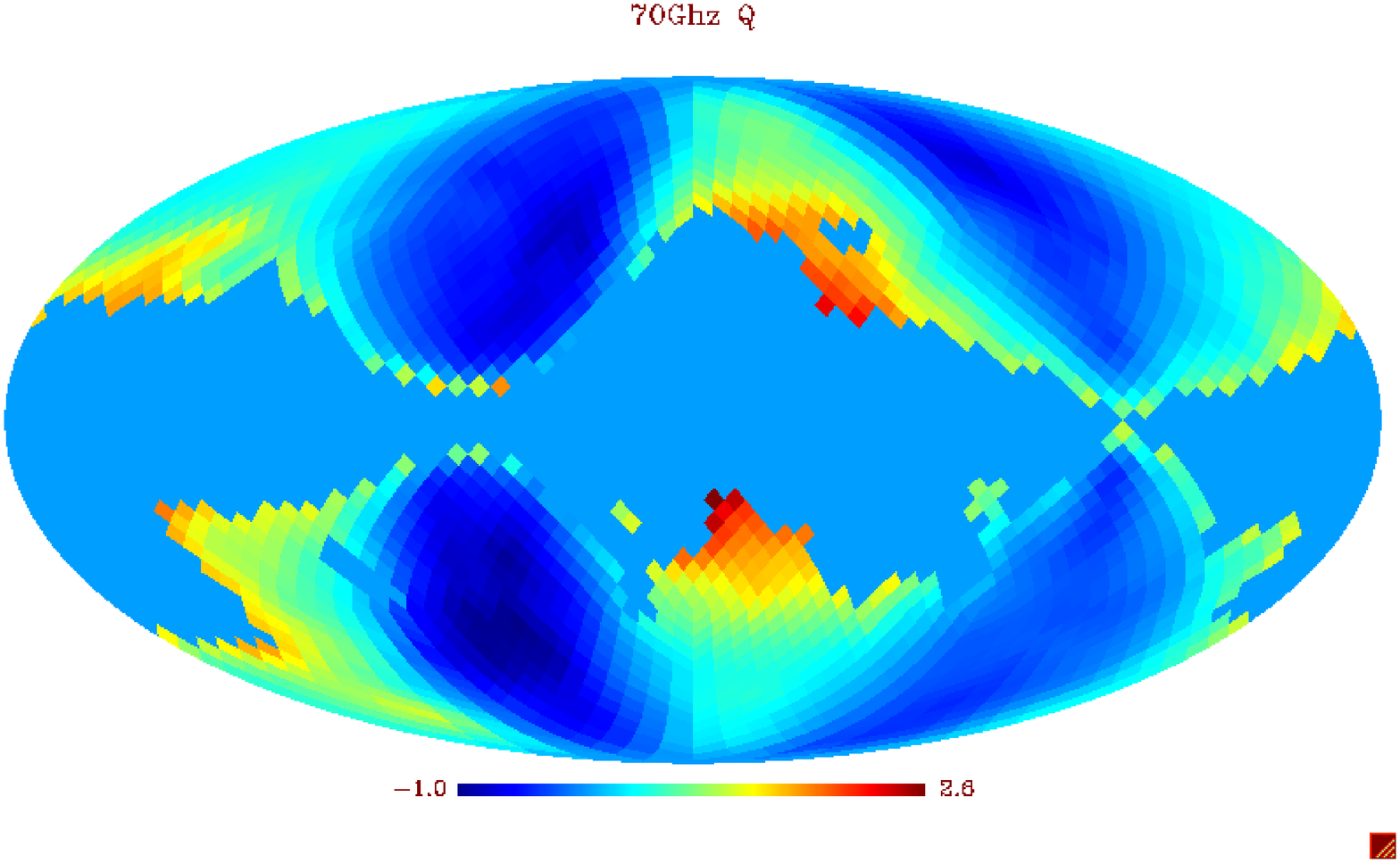}
\includegraphics{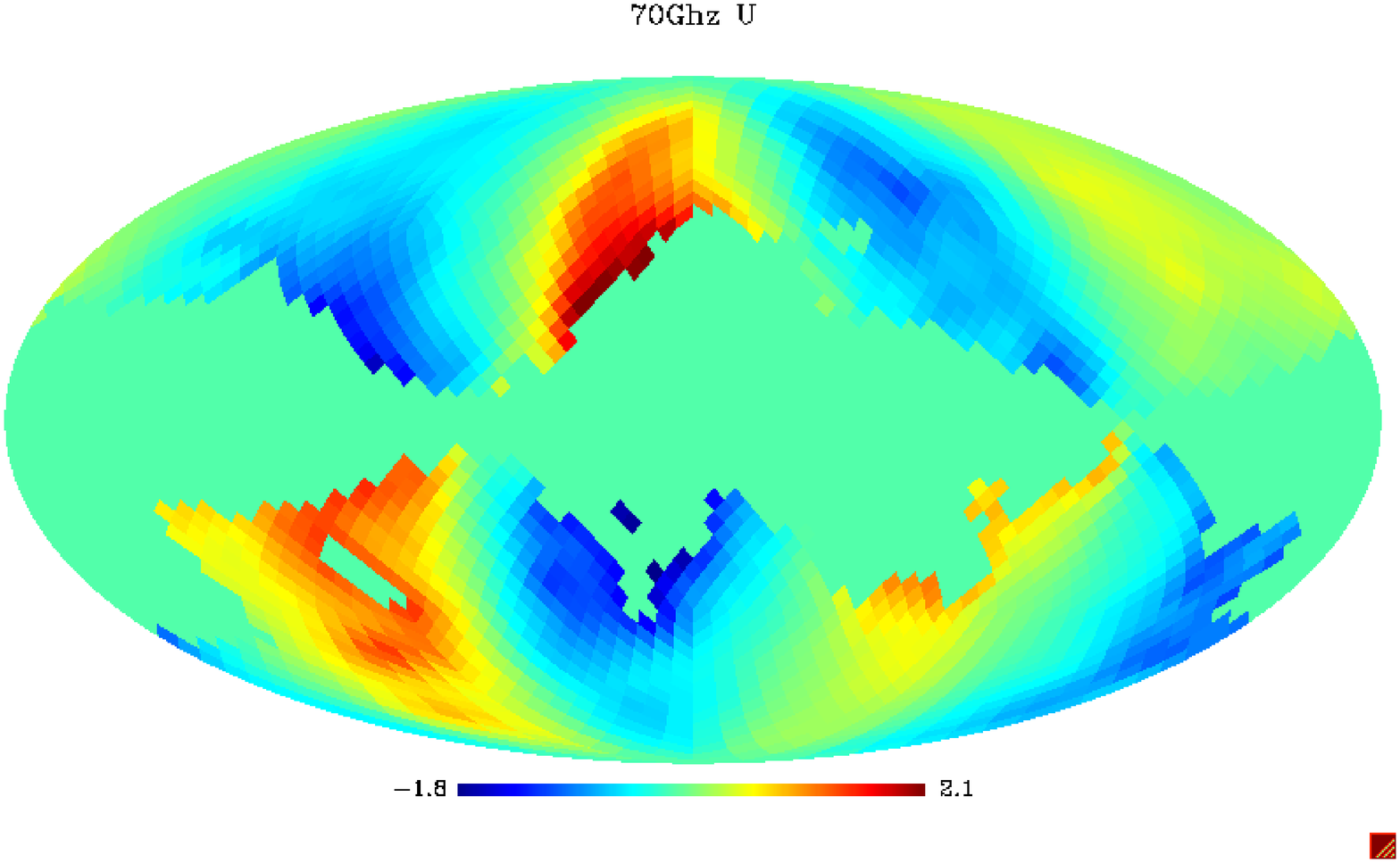}
\includegraphics{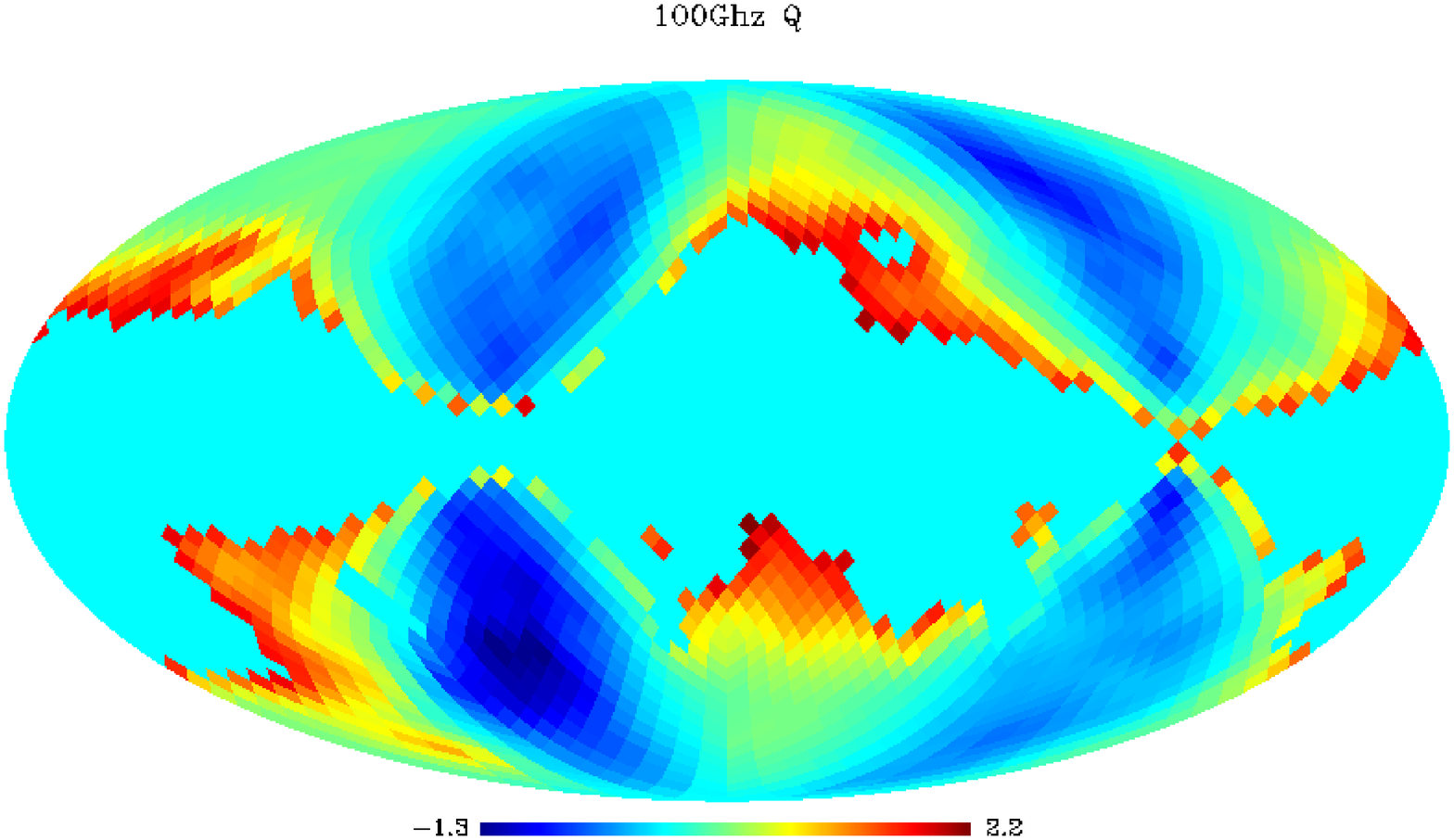}
\includegraphics{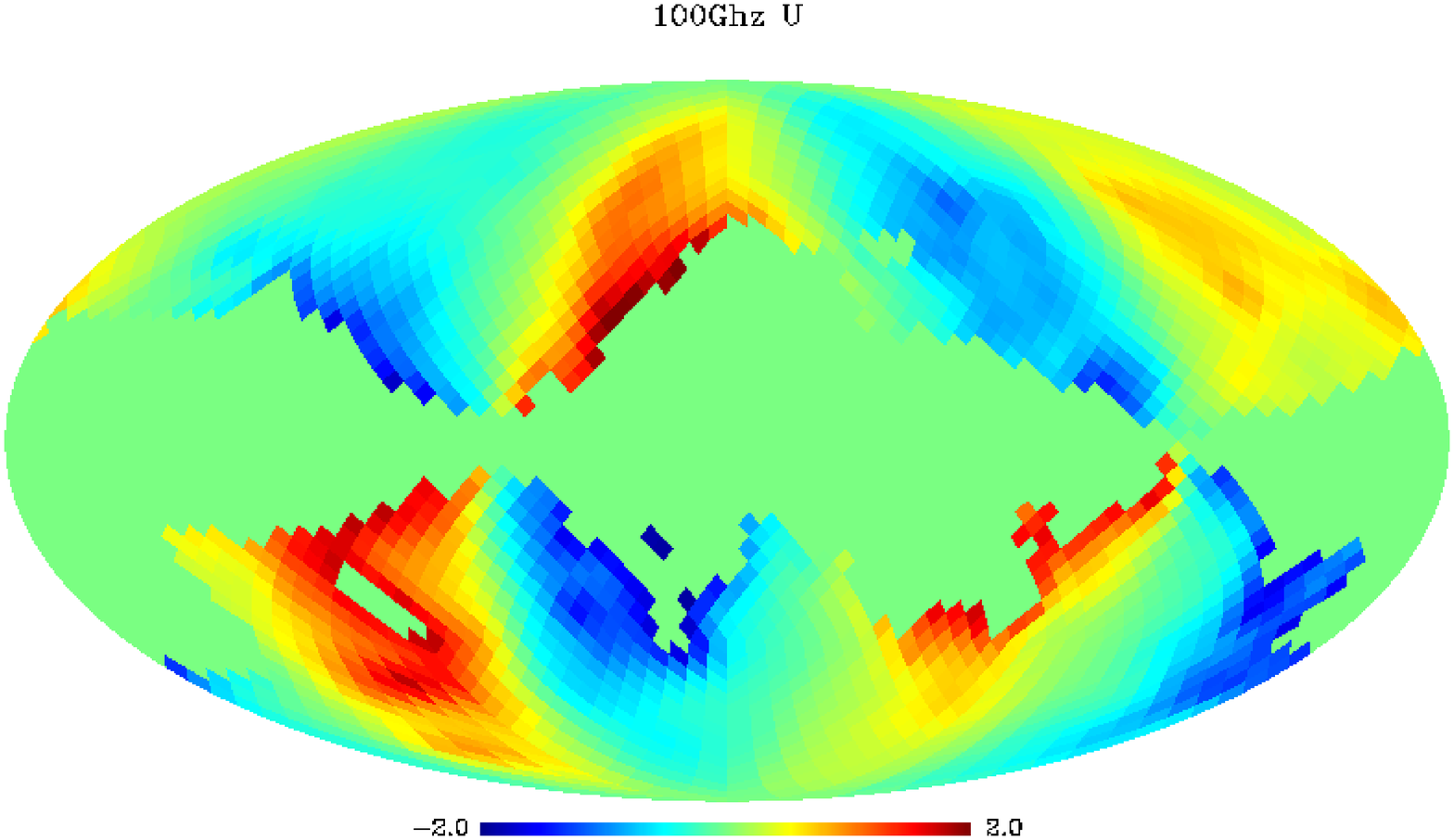}
\includegraphics{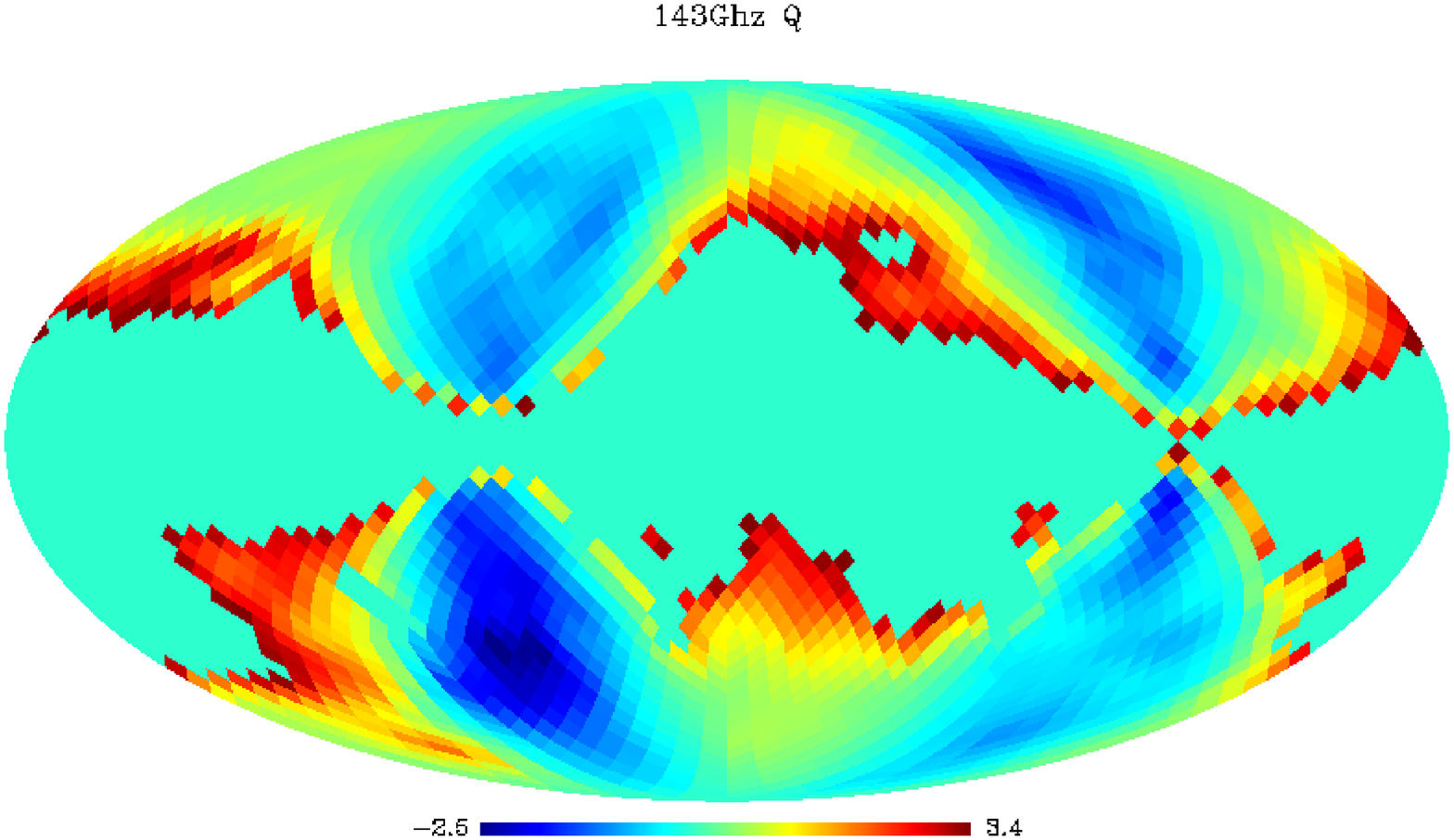}
\includegraphics{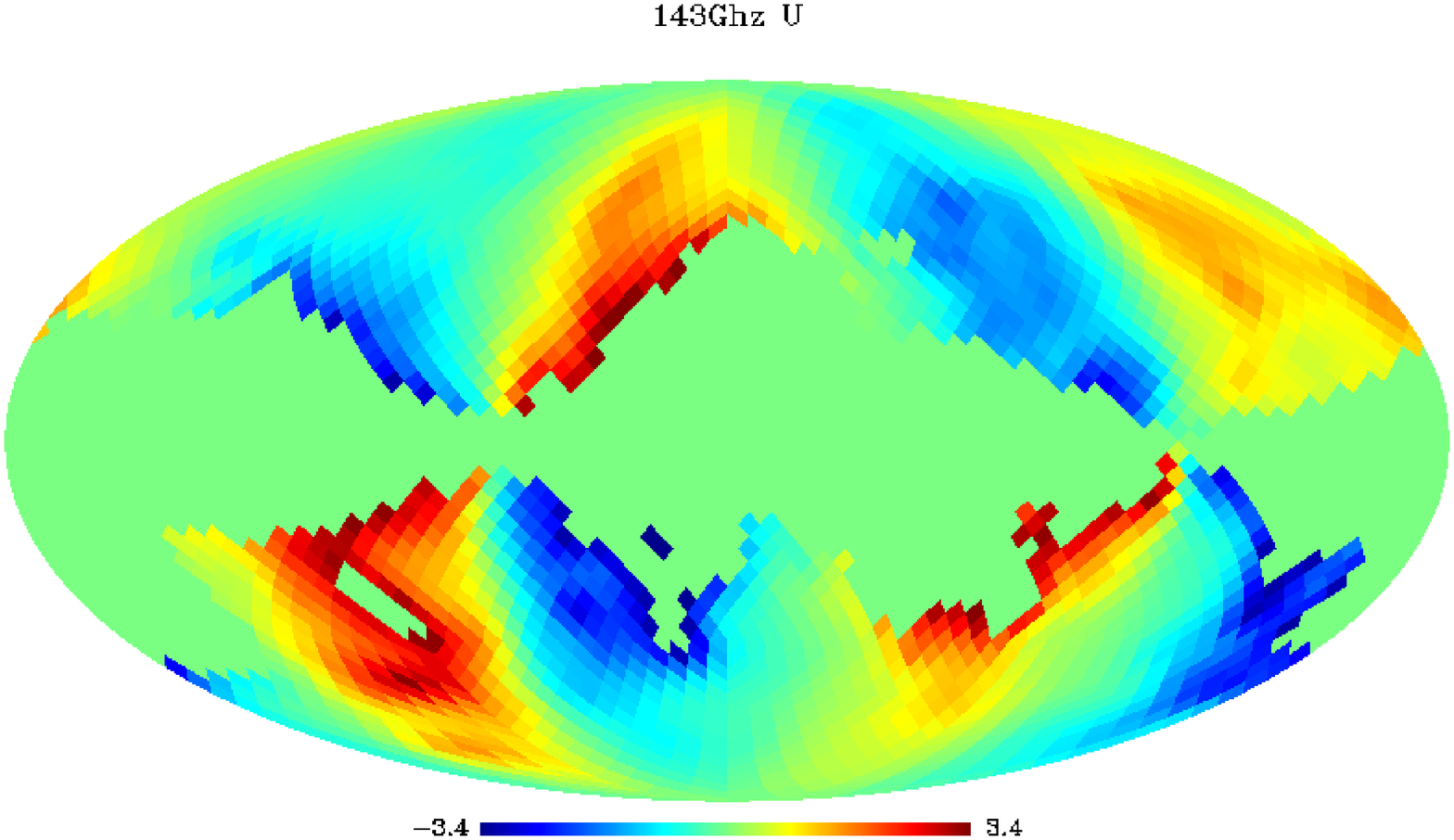}
\includegraphics{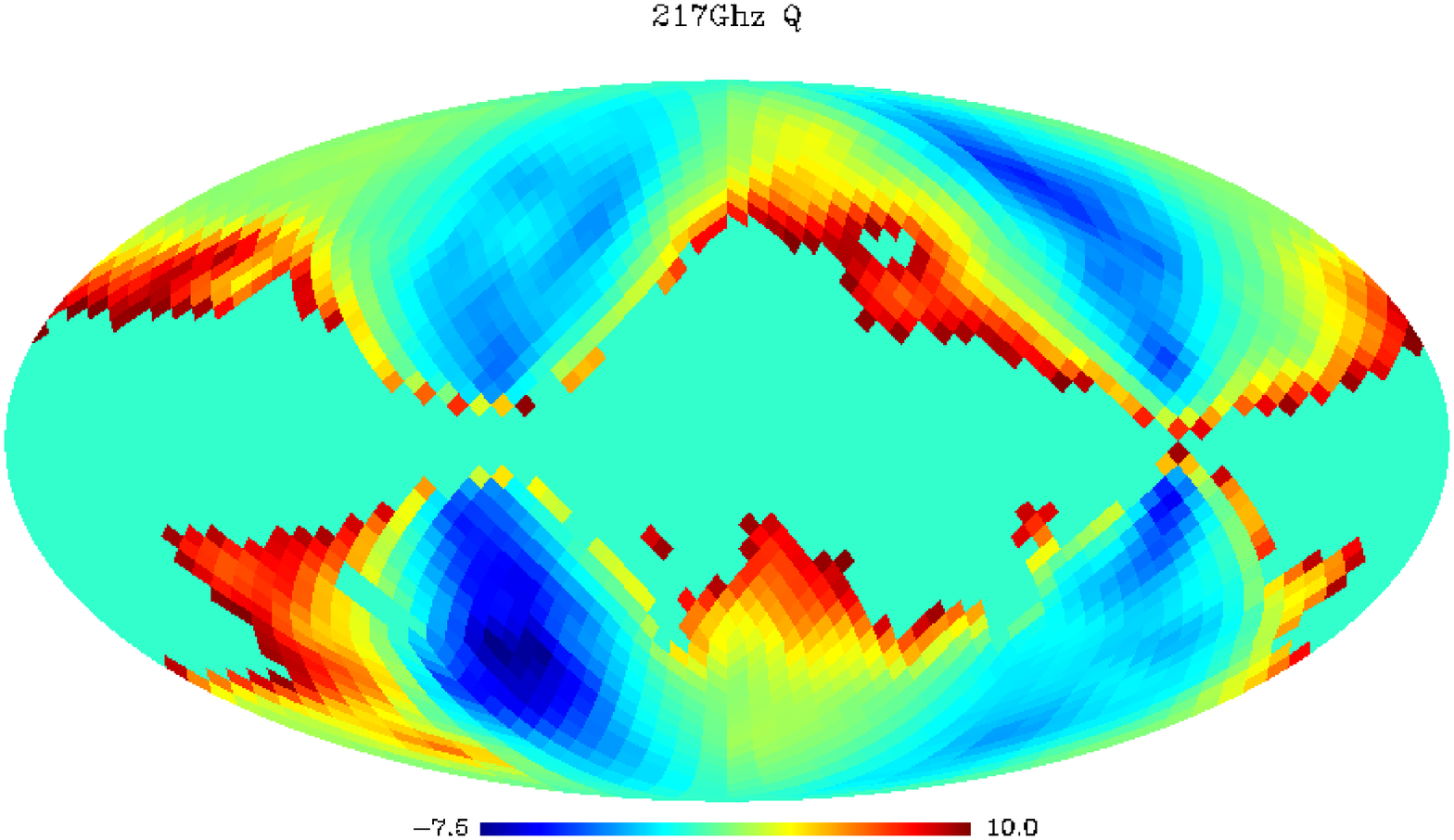}
\includegraphics{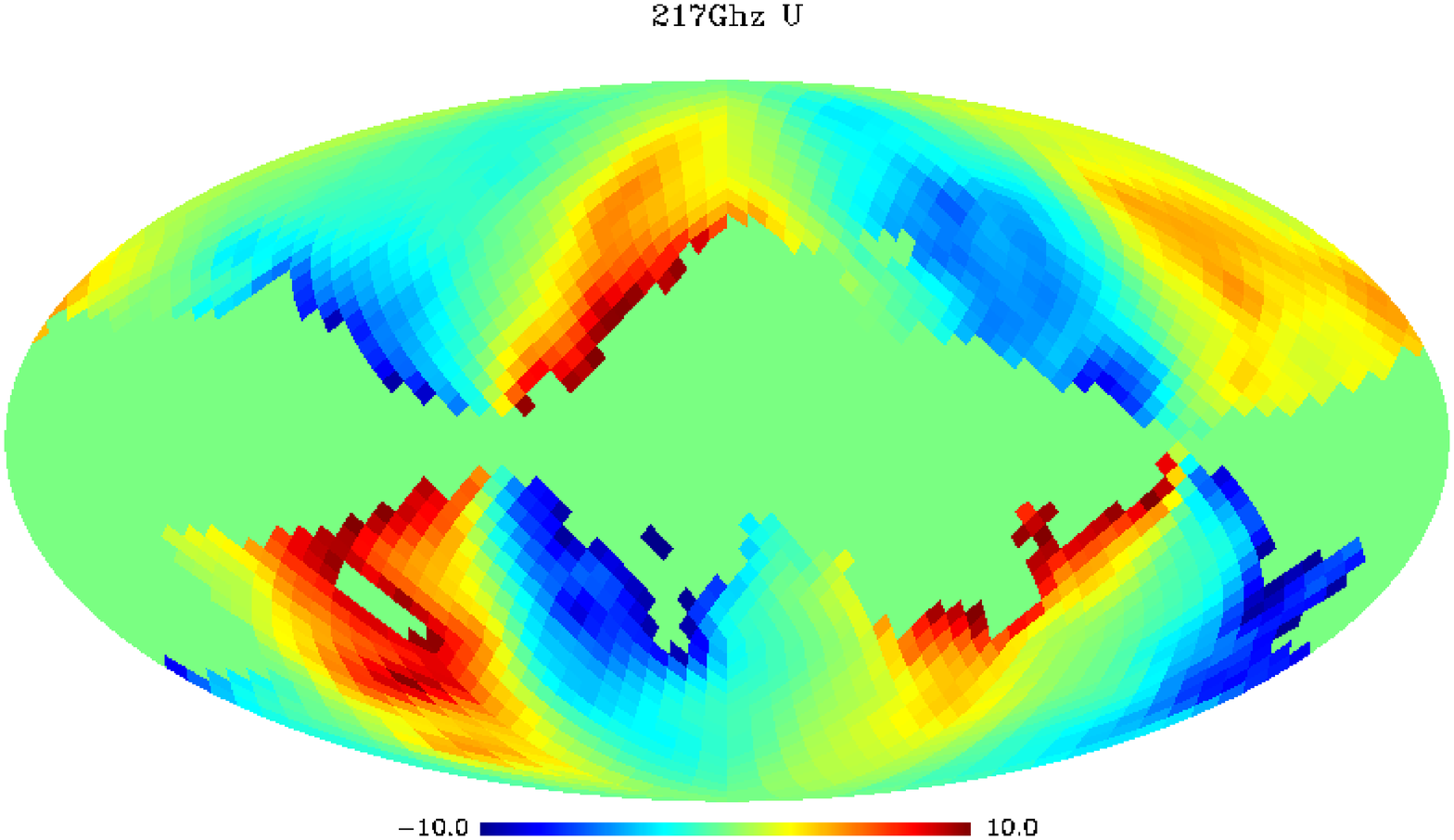}

\caption
{As Figure 1, but with maps generated at Healpix NSIDE=16 and a smoothing of $7^\circ$ FWHM.
The internal mask described in the text has been applied to the PSM.}

\label{figure2}
\end{figure*}

Figure 1 shows Q and U simulations at Healpix (Gorski \etal, 2005) resolution
NSIDE=2048 for a single realization of the concordance cold dark
matter model\footnote{Throughout this paper, apart from the
  tensor-scalar ratio $r$, we use the cosmological parameters derived
  from the 3-year WMAP data (Spergel \etals 2007) assuming zero
  curvature and a single scalar spectral index.  The tensor spectral
  index is fixed at $n_t = 1$.}.  The model has a tensor-scalar ratio
$r=0.1$. The PSM of polarized foregrounds is shown in the lower panels over the
frequency range $70$-$217$ GHz with the colour (grey) scales adjusted
to span the range $-30\; \mu$K to $30 \; \mu$K. The structure of the
foreground is fairly similar at each of these frequencies, and so we
use the $217$ GHz maps to define a polarization mask by simply applying
a threshold to each of the $Q$ and $U$ maps. An `internal mask' is
then defined as the union of the two $Q$ and $U$ masks. For the tests
described in this paper we use a fairly conservative mask which
removes $37\%$ of the sky.

\subsection{Impact of foregrounds on B-mode detection}

Figure 1 is not particularly useful for assessing tensor mode
detection with \planck, since \plancks will be noise dominated in the
$B$-mode for all but the lowest multipoles. Smoothed maps, as shown in
Figure 2, are of more relevance. Here we show the maps of Figure 1 in
the regions outside the internal mask after smoothing with a Gaussian
of FWHM $7^\circ$ and repixelization to a resolution of NSIDE=16. One
can see that at this resolution (almost signal dominated for Planck)
the peak-to-peak variations in the polarization maps of the primary
CMB are of order $\sim 0.7$ $\mu$K.  The PSM is shown for the regions
outside the internal mask, but unlike Figure 1, the temperature scales
of the colour tables are set by the true maximum and minimum values of
the maps. Evidently at the most sensitive \plancks channel ($143$ GHz)
foregrounds dominate over the primary CMB signal over the whole
sky. In fact, to get a better feel for how accurately we need to
subtract foregrounds for $B$-mode detection, Figure 3 shows the
contributions to the $Q$ and $U$ maps from the $B$-mode alone for
$r=0.1$. The {\it rms} contribution from the $B$-mode at this
resolution is about a quarter of the {\it rms} of the $E$-mode, and so
foreground subtraction to significantly better than $5\%$ accuracy at
$143$ GHz is required to detect a $B$-mode with $r=0.1$. As we will
show, this presents a formidable challenge for \planck, even with
the simplified foreground model assumed in the PSM.  The {\it rms}
values for the PSM and for the primary CMB maps shown in Figures 2 and
3 are listed in Table 1. Power-spectra for the maps shown in Figure 2,
computed using the pseudo-$C_\ell$ (PCL) estimator\footnote {Note that as
  described in Efstathiou (2006) these PCL estimates are highly
  sub-optimal at low multipoles, but they are perfectly adequate for
  illustrating the magnitude of the foreground problem. We will
  discuss more optimal methods in Sections 3 and 4.},
are plotted in Figure 4.

\begin{figure*}
\vskip 2.5 truein

\includegraphics{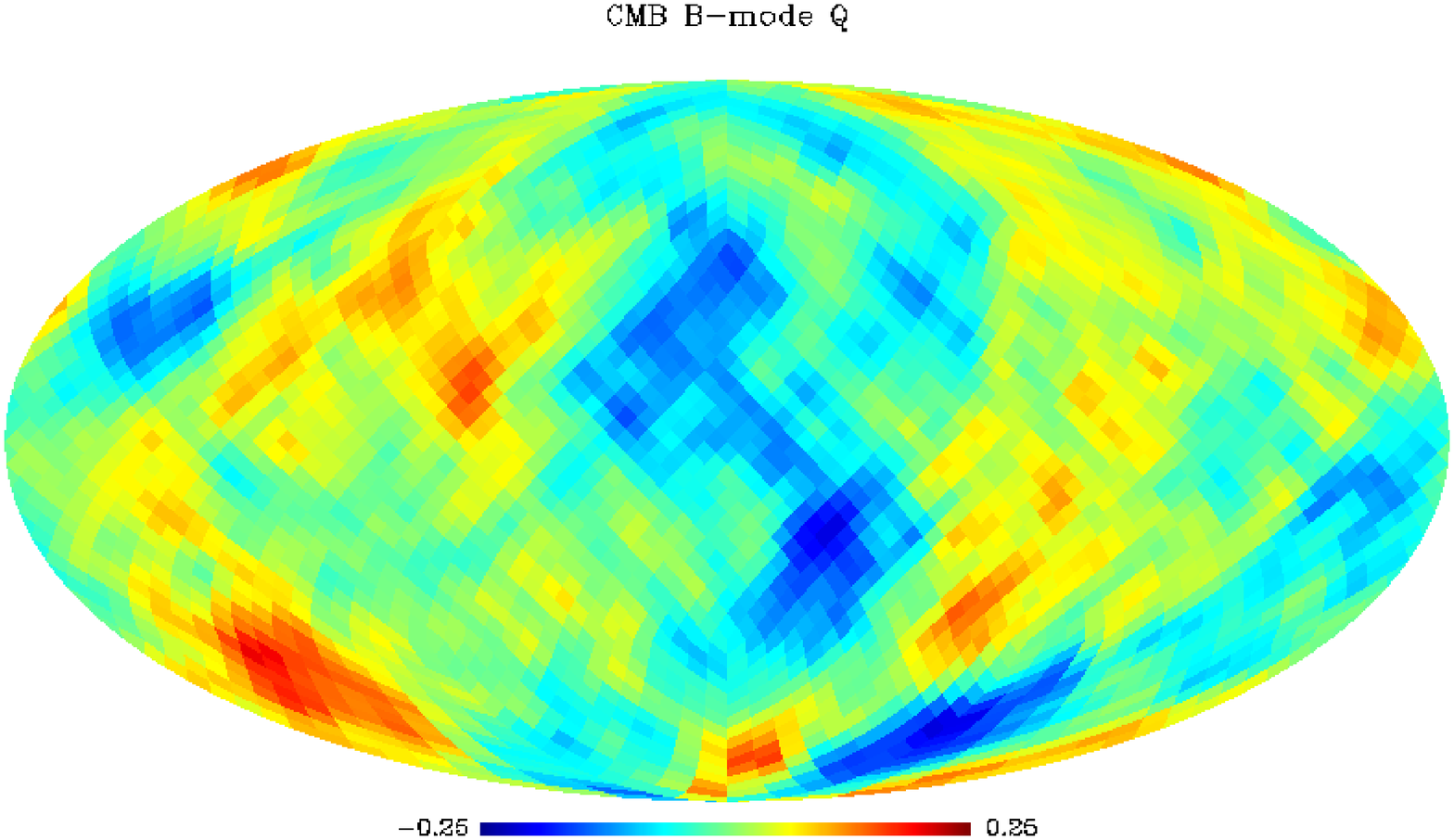}
\includegraphics{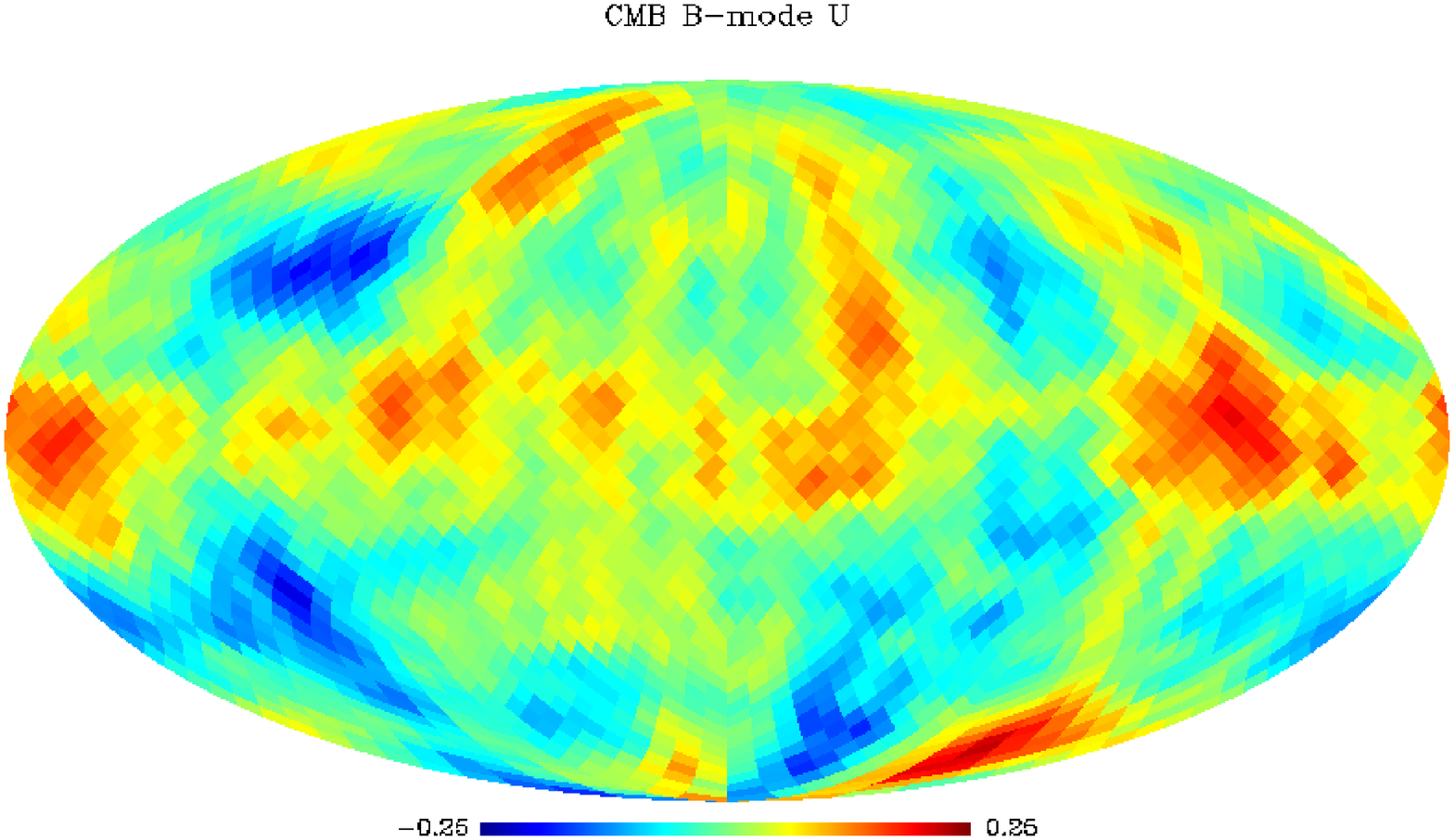}

\caption
{Maps of the $B$-mode contributions to the primary CMB realizations ($r=0.1$) shown in Figure 2.}

\label{figure3}
\end{figure*}

\begin{table}

\centerline{\bf \ \ \  Table 1:  RMS residuals outside internal mask$^1$}

\begin{center}

\begin{tabular}{cccc} \hline \hline
\smallskip 
Map       & $T$ ($\mu {\rm K}$) & $Q$ ($\mu {\rm K}$) & $U$ ($\mu {\rm K}$) \cr
\hline \hline
CMB  & $50.4$ & $0.257$  & $0.232$   \cr
CMB B-mode    & $2.26$ &  $0.062$ & $0.064$ \cr
PSM $\;\;30$ GHz  & $97.2$ &  $4.29$ & $3.94$ \cr
PSM $\;\;44$ GHz  & $36.0$ &  $1.51$ & $1.37$ \cr
PSM $\;\;70$ GHz  & $19.0$ &  $0.661$ & $0.608$ \cr
PSM $100$ GHz  & $22.7$ &  $0.720$ & $0.692$ \cr
PSM $143$ GHz  & $40.8$ &  $1.29$ & $1.27$ \cr
PSM $217$ GHz  & $119.2$ &  $3.82$ & $3.76$ \cr
PSM $353$ GHz  & $874.1$ &  $28.3$ & $27.9$ \cr
\hline
\end{tabular}

\end{center}
\begin{center}
$\;$$^1$  For maps smoothed with a Gaussian of FWHM $7^\circ$.
\end{center}
\end{table}

\begin{figure*}
\vskip 3.5 truein

\includegraphics{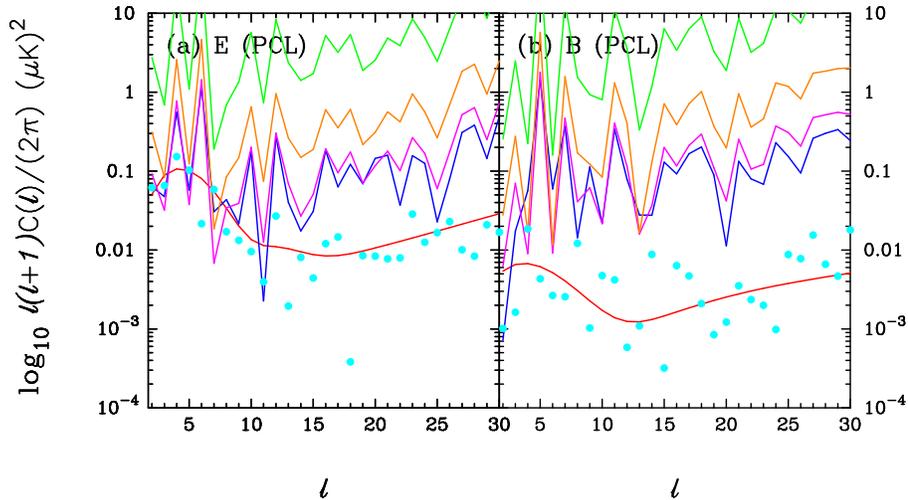}

\caption {PCL $E$ and $B$-mode power spectrum estimates computed for
  the CMB simulations and foreground components of
  Figure 1. The power spectra are computed for the region of the sky
  outside the internal mask. No instrumental noise has been added to
  the simulations.  The blue points show the power spectrum estimates
  for the CMB.  The red lines show the theoretical input CMB
  spectra. The foreground power spectra are as follows: 70 GHz (dark
  blue); 100 GHz (purple); 143 GHz (orange); 217 GHz (green).}

\label{figure4}
\end{figure*}

\section{ILC Component Separation}

\subsection{Summary}
The internal linear combination method is very simple and well known
(see {\it e.g.} Bennett \etals 2003; Eriksen \etals 2004b). Suppose we
have $M$ maps $T^i(p)$ (temparature, $Q$, or $U$ polarization at each
pixel $p$) at different frequencies, we find the linear combination
\begin{equation}
  T(p) = \sum_i^M w_i T^i(p) \label{ilc1}
\end{equation}
that minimises the variance
\begin{equation}
 {\rm Var}(T) = \langle T^2 \rangle - \langle T \rangle^2  \label{ilc2}
\end{equation}
subject to the constraint that the sum over the $w_i$ is equal
to unity. (Angular brackets in this Section denote averages over map pixels.) 
Equation (\ref{ilc2}) can be minimised using a Lagrange
multiplier.  If the map covariance matrix
\begin{equation}
M_{ij} = {1 \over N_p} \sum_p (T^i(p) - \langle T^i \rangle)(T^j(p) - \langle T^j \rangle )
 \label{ilc4} 
\end{equation}
is invertible then the solution is
\begin{equation}
    w_i = {\sum_k M_{ki}^{-1} \over \sum_{kj} M_{kj}^{-1}}. \label{ilc3}
\end{equation}
and we denote the corresponding map  $T^{\rm ILC}(p)$.  (If $M_{ij}$ is not invertible then a family of solutions for $w_i$ exists.)
As a slight variation of this method for noisy data,
one might consider subtracting the noise variance from equation (\ref{ilc4}) so that
we minimise primarily on the foreground residuals. In this case, the
solution is identical to (\ref{ilc3}) with $M_{i j}$ replaced by
\begin{equation}
 M_{ij} \rightarrow  M_{ij} -   \overline \sigma^2_j  \delta_{ij}, \label{ilc5a}
\end{equation}
where $\overline \sigma^2_i$ is the noise variance of the map at
frequency $i$. (Note that this variant is inconsistent with a Bayesian
formulation of ILC presented in Gratton (2008).)

\subsection{ILC with variable foregrounds}

Let us ignore instrumental noise, but assume that the foregrounds
differ with frequency. At each frequency $i$, we can write the data
vector as
\begin{equation}
  T^i  = S + F^i \label{ilc5}
\end{equation}
where $S$ is the frequency independent CMB signal and $F^i$ is the
foreground. (For simplicity, we assume that $S$ and $F^i$ have zero
mean). To find the ILC solution, we must extremize the quantity
\begin{equation}
  \Sigma^2  = \sum_{ij} w_i w_j \langle (S  + F^i)(S + F^j) \rangle + \lambda \left (\sum_i w_i - 1 \right) \label{ilc6}
\end{equation}
with respect to the weights $w_i$ and $\lambda$. The solution is
\begin{equation}
  w_i = { \left ( 1 + \sum_{kj} X_k F_{kj}^{-1} \right ) \sum_{k} F_{ki}^{-1}
\over \sum_{kj} F_{kj}^{-1}} - \sum_k X_k F_{ki}^{-1}, \label{ilc7}
\end{equation}
where 
\begin{equation}
\left. \begin{array}{ll}
 X_i & = \langle SF^i \rangle, \\
 F_{ij} & = \langle F^iF^j \rangle, 
\end{array} \right \}    \label{ilc8}
\end{equation}
and we have assumed in (\ref{ilc7}) that there is enough variation in
the foregrounds with frequency that the matrix $F_{ij}$ is
non-singular and therefore invertible.  Notice that in the limit of
zero signal, $X_i=0$, the solution is just equation (\ref{ilc3}) with
$M_{ij} = F_{ij}$. In this case the variance of the ILC solution is
\begin{equation}
\langle (\Delta T^{\rm ILC})^2 \rangle =  \left ( \sum_{ij} F_{ij}^{-1} \right )^{-1}. \label{ilc8a}
\end{equation}
This gives an indication of the best possible foreground subtraction
achievable through ILC, {\it i.e.} of the limitation imposed by {\it
  `foreground mismatch'}. (We will apply the term `foreground
mismatch' generally to subtraction techniques to denote the residual
contamination by foregrounds even in the absence of CMB signal and
instrument noise.)  The terms in equation (\ref{ilc7}) proportional to
$X_i$ lead to an offset in the ILC solution with an amplitude
proportional to $X_i$ and {\it independent of the amplitude of the
  foreground}.  Such offsets exist even if there is not enough
variation in the foregrounds to mimic the CMB.  Consider the
well known case of two frequency channels with an identical
foreground template:
\begin{equation}
\left. \begin{array}{ll}
 T_1 & = S + \alpha F, \\
 T_2 & = S + F, 
\end{array} \right \}    \label{ilc9}
\end{equation}
(without loss of generality, since there are only two unknown
functions, $S$ and $F$) . The ILC solution in this case gives
\begin{equation}
  T^{\rm ILC} = S  - {\langle SF \rangle  \over \langle F^2 \rangle} F ,  \label{ilc10}
\end{equation}
independent of the parameter $\alpha$. The error in (\ref{ilc10}),
which we will term `{\it cross-correlation offset}' (called `cosmic
covariance' by Chiang, Naselsky and Coles, 2008), is independent of the
amplitude of the foreground template and is irremovable, since the
amplitude of the effect depends on the specific realization of the CMB
signal.  Equation (\ref{ilc8}) generalises this result to an arbitrary
number of channels and includes the bias caused by foreground
mismatch.  The cross-correlation offset averages to zero over a large
number of realizations of the CMB. However, it can be significant for
any single realization.  Cross-correlation offset is particularly
serious for B-mode polarization measurements. This is because we are
trying to detect a $B$-mode signal that is a small fraction of the
$E$-mode contribution to the $Q$ and $U$ maps ({\it cf}.\ Figure 3).
The cross-correlation terms $X_i$ for the $Q$ and $U$ maps are
therefore dominated by the contribution from $E$-modes, and since
these offset terms are fixed for a particular realization of the
$E$-modes, they set a fundamental limit on the amplitude of a
  $B$-mode signal detectable via ILC subtraction.

\begin{figure*}
\vskip 3.2 truein

\includegraphics{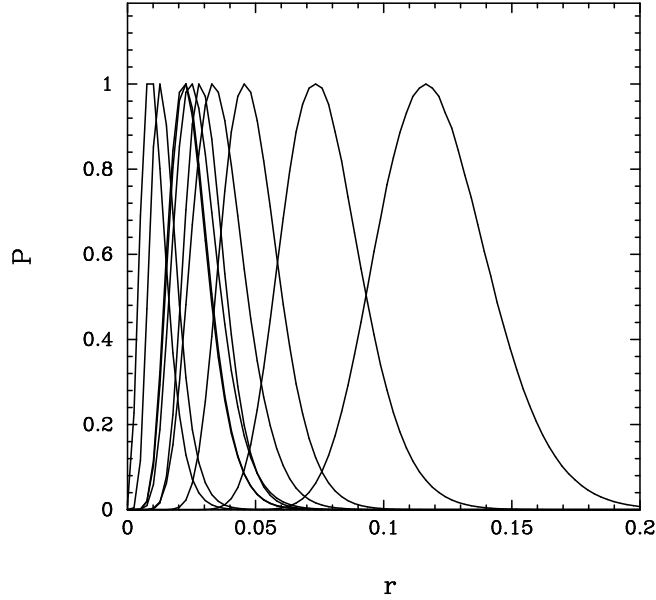}

\caption
{Distributions of the tensor-scalar ratio $r$ for 10 simulations
with ILC cleaning as described in the text. The simulations were
generated with $r=0$.}

\label{figlike}
\end{figure*}

This is illustrated by Figure \ref{figlike}, which shows the likelihood
distributions of the tensor-scalar ratio, $r$, for ten ILC cleaned CMB
realizations generated with $r=0$. In each case, we generated an
NSIDE=16 low resolution CMB map smoothed with a Gaussian of FWHM
$7^\circ$ and added the PSM foregrounds at the four frequencies $70$,
$100$, $143$ and $217$ GHz as plotted in Figure 2.  The internal mask
of Figure 2 was applied to each map and an ILC cleaned map was
computed from the solution (\ref{ilc3}). 

The pixel likelihood function is
\begin{equation}
{\cal L} \propto {1 \over \sqrt{ \vert {\bf C} \vert}} {\rm exp} \left 
( - {1 \over 2} {\bf x}^T{\bf C}^{-1} {\bf x} \right ) ,   \label{like1}
\end{equation}
where ${\bf x}$ is the $(T, Q, U)$ ILC map and ${\bf C}$ is the sum of the signal 
(${\bf S}$) and noise (${\bf N}$) covariance matrices
\begin{equation}
C_{ij} = \langle x_ix_j \rangle, \quad {\bf C} = {\bf S} + {\bf N}. \label{like2}
\end{equation}
In the examples shown in Figure \ref{figlike}, we compute the likelihood
function for the $Q$ and $U$ maps allowing only the parameter $r$ to vary.
(This is a very good approximation to a full likelihood analysis because
$r$ is weakly correlated with other cosmological parameters.) A small
diagonal noise of $0.1$ $\mu$K was added to the $Q$ and $U$ maps to regularise
the signal covariance matrix, which is otherwise numerically singular
because the maps are over-pixelized. Figure \ref{figlike} shows that the
cross-correlation offset causes biases $r \sim 0.03$, but with 
a large spread so that biases in excess of $r \sim 0.10$ are seen. 
The results shown in Figure 5 are insensitive to the number of channels
used in the ILC.

\begin{figure*}
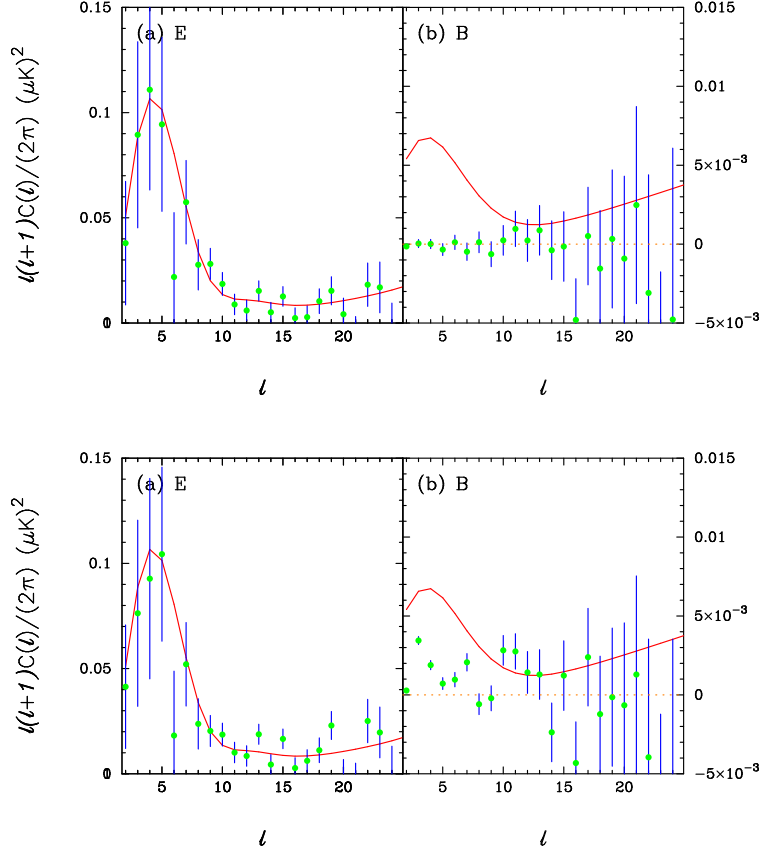

\vskip 4.5
 truein

\includegraphics{pgqml1.ps}

\includegraphics{pgqml2.ps}

\caption {QML estimates of the $E$ and $B$-mode polarization spectra
  for the realization plotted in Figure 3 with $r=0$. The upper panel
  shows the analysis of the CMB maps on the cut sky. The lower panels
  show the analysis of the ILC cleaned CMB maps.  The error bars show
  the diagonal components of (\ref{ML4}) using the theoretical input
  values of $r$ for each realization. The solid lines show the
  theoretical input spectra for $r=0.1$.}
\label{figqml}
\end{figure*}

The effects of the cross-correlation offset on the power spectra
are illustrated in Figure \ref{figqml} for the specific
CMB realization of Figure 2 with $r=0$. Here we plot the 
quadratic maximum likelihood (QML) estimates  (Tegmark and de Oliveira-Costa
2001; Efstathiou 2006)
\begin{equation}
\hat C_{\ell} = {\cal F}^{-1} y,   \label{ML1}
\end{equation}
where 
\begin{equation}
 y^r_{\ell} = x_i x_j E^{r\ell}_{ij},  \qquad r \equiv (T, X, E, B),   \label{ML2a}
\end{equation}
with matrices $E^{r\ell}$ given by 
\begin{equation}
 E^{r \ell} = 
{1 \over 2}C^{-1} {\partial C \over \partial C^r_\ell} C^{-1},  \label{ML2b}
\end{equation}
$x$ and $C$ are the data vector  and data covariance matrix defined
in equation (\ref{like1}), and ${\cal F}$ is the Fisher matrix
\begin{equation}
{\cal F}^{mn}_{\ell \ell^\prime} = {1 \over 2}
{\rm Tr} \left [  {\partial C \over \partial  C^m_{\ell^\prime}}  C^{-1}
{\partial C \over \partial  C^n_{\ell}}  C^{-1} \right ]. \label{ML3}
\end{equation}
The covariance matrix of the QML estimates is given by the 
inverse of the Fisher matrix (\ref{ML3})
 \begin{equation}
 \langle \Delta \hat C_{\ell} \Delta \hat C_{\ell^\prime} \rangle  = {\cal F}^{-1}.  \label{ML4}
\end{equation}
The upper panels of Figure \ref{figqml} show the QML estimates for the
CMB maps alone computed from the cut sky. (Regularizing noise of
$0.1$ $\mu$K was added to the maps as in the likelihood analysis of
Figure \ref{figlike}.) The error bars on the points are computed from
the diagonal components of the Fisher matrix ({\ref{ML3}) assuming
$C^B_\ell=0$.  The lower panels show the QML power spectra for the
ILC cleaned maps, illustrating the cross-correlation offset.

A significant cross-correlation offset is inherent in any `blind'
component separation technique. For the ILC method as presented above,
the size of the offset is set by the amplitude of the $E$-mode signal
with a resulting large effect on the $B$-mode. In principle, the
cross-correlation offset could be reduced by first disentangling $E$
and $B$-modes and performing ILC independently on maps constructed
from the two sets of modes. If the whole sky is available, the
separation into $E$ and $B$-modes is exact and unambiguous. If the sky
coverage is incomplete, one could decompose the maps into almost pure
$E$ and $B$-modes following the techniques described by Lewis
(2003). Such a decomposition will necessarily lead to information loss
since ambiguous modes must be discarded.  Even if near pure $B$-modes
are identified, so reducing the cross-correlation offset, we will see
in the next subsection that an ILC solution can amplify the
instrument noise to unacceptably high levels. As we will show in
Section 4, rather than invoke some type of modal decomposition, it is
possible to reduce the cross-correlation offset to negligible levels by
template fitting and to avoid catastrophic amplification of instrument
noise. The scheme outlined in Section 4 has the added advantage of
leading to a simple model for the full polarization likelihood
function.

\subsection{ILC with noise}

So far, we have ignored instrumental noise. In the presence of
instrumental noise, the solution for the ILC weights is approximately
(\ref{ilc7}) with $F_{ij}$ replaced by
\begin{equation}
\tilde F_{ij}  = F_{ij}+ \bar \sigma^2_i \delta_{ij},  \label{ilcn1}
\end{equation}
where $\bar \sigma^2_{i}$ is the noise variance of the map at
frequency $i$. (This solution is exact if the noise-signal $\langle NS
\rangle$ and noise-foreground $\langle NF \rangle$ covariances can be
neglected.) If noise dominates in the frequency covariance matrix, the
ILC solution corresponds to inverse noise-variance weighting,
\begin{equation}
w_i = {1 \over \bar \sigma^2_i} \left ( \sum_k {1 \over \bar \sigma^2_k} \right )^{-1}, \label{ilcn2}
\end{equation}
which has, of course, nothing to do with foregrounds. One can get a rough 
idea of how noise will affect the ILC solution by computing 
\begin{equation}
\langle (\Delta T^{\rm ILC})^2 \rangle =  \left ( \sum_{ij} \tilde 
F_{ij}^{-1} \right )^{-1} \label{ilcn3}
\end{equation}
instead of (\ref{ilc8}). In the limit that noise dominates, the
ILC residuals will simply reflect the noise level of the
ILC map,
\begin{equation}
\bar \sigma_{\rm ILC}^2 = \sum_i w^2_i \bar \sigma^2_i
\approx \left ( \sum_i {1 \over \bar \sigma^2_i } \right )^{-1}, \label{ilcn4}
\end{equation}
where we have assumed that the noise at each frequency is  independent.

\begin{table}

\centerline{\bf \ \ \  Table 2:  ILC weights}

\begin{center}

  \begin{tabular}{ccccccc} \hline \hline \smallskip & $70$ GHz &
    $100$ GHz & $143$ GHz & $217$ GHz & & \cr Galaxy no noise & $w_1$ &
    $w_2$ & $w_3$ & $w_4$ & $\langle (T^{\rm ILC} - T^{\rm CMB})^2
    \rangle^{1/2}$ ($\mu$K) \cr \hline\hline T &
    $-0.1427$ & $-1.0457$ & $\;\;\;2.995$ & $-0.8070$& $0.947$ \cr Q &
    $-1.5525$ & $\;\;\;5.2210$ & $-2.936$ & $\;\;\;0.2675$ & $0.0054$ \cr U &
    $-1.6387$ & $\;\;\;5.5929$ & $-3.285$ & $\;\;\;0.3303$ & $0.0047$ \cr \hline

 Galaxy+CMB & $w_1$ & $w_2$ & $w_3$ & $w_4$ & $\langle (T^{\rm ILC} - T^{\rm CMB})^2
    \rangle^{1/2}$ ($\mu$K)  \cr \hline\hline 
T & $\;\;\;37.191$ & $-129.68$ & $113.52$ & $-20.028$ & $9.547$
    \cr 
Q & $-5.3553$ & $\;\;\;\;\;19.526$ & $-15.707$ & $\;\;\;\;2.536$ & $0.0578$ \cr 
U & $-6.7704$ & $\;\;\;\;\;26.042$ & $-21.886$ & $\;\;\;\;3.614$ &  $0.0619$ \cr \hline
\end{tabular}
\end{center}
\noindent
Notes:  The upper table lists the ILC weights, $w_i$, computed from the
foregrounds alone ({\it i.e.} equation (\ref{ilc3}) with 
$F_{ij}$ replacing $M_{ij}$). The column labelled 
$\langle (T^{\rm ILC} - T^{\rm CMB})^2 \rangle^{1/2}$ lists the
rms residual of the ILC cleaned maps and the true CMB maps for the
regions outside the internal mask shown in Figure 2. 
The lower table lists the ILC weights for the combined
noise-free foreground and CMB maps shown in Figure 2. 
\end{table}

Table 2 lists the ILC weights for the foregrounds alone (equation
\ref{ilc3} with $F_{ij}$ replacing $M_{ij}$) computed for the pixels
outside the internal mask shown in Figure 2. The ILC residuals using
these weights are listed in the last column of the table. For the Q
and U maps the residuals are an order of magnitude lower than the {\it
  rms} contribution expected for a B-mode with $r=0.1$. Thus, if the
PSM model is correct, it is in principle possible to assign weights
that subtract foregrounds to achieve a limit of $r \sim 10^{-3}$ ({\it
  i.e.} for the PSM, `foreground mismatch' is negligible).  However,
three of the four weights are considerably larger than unity and will
therefore amplify any instrument noise that is present (equation
\ref{ilcn4}). If somehow we were given these weights, we would only
able to make use of them if the instrument noise at each channel were
extremely low. For example, using the instrumental sensitivities in SPP05
for \planck, these weights would lead to disastrously high noise
levels of $\sim 0.6$ $ \mu{\rm K}$ for the ILC cleaned $Q$ and $U$ maps. Such
a high noise level is of no use for B-mode detection.

The situation become even worse if we include the CMB in the ILC
solution (lower entries in Table 2). The ILC solution listed in Table
2 is for the CMB realization of Figure 2 with $r=0$. The polarization
power spectra for this solution are shown in the lower panel of Figure
\ref{figqml}.  The cross-correlation offset amplifies the ILC
residuals by more than an order of magnitude compared to the
foreground-only solution (see the last column in Table 2) and so some
foreground $B$-mode leaks into the power spectrum at low
multipoles as shown in Figure 6. This is consistent with the fundamental limit of $r \sim
0.1$ imposed by the cross-correlation offset discussed in the previous
section. However, notice that the weights are now even larger than for
the foreground-only case and so any instrumental noise will be highly
amplified in the ILC solution.

For realistic experiments, we are therefore caught between a rock and
a hard place.  In the presence of instrumental noise, we would like to
minimise the noise when combining frequency channels (equation
\ref{ilcn2}). However, this will not remove foregrounds. If we use
weights that minimise the foreground residuals (which we cannot find
in principle because of cross-correlation offset) we amplify the noise
to unacceptable high levels.

\section{Template Fitting}

The discussion of the previous Section shows that a purely blind
component separation method such as ILC is fundamentally limited for
$B$-mode detection by the cross-correlation offset, even if foreground
mismatch is negligible ({\it i.e.}  a linear combination exists which
eliminates the foregrounds to high accuracy).  To reduce
the cross-correlation offset, a semi-blind technique is required which
utilises supplementary information on either the foregrounds or the
primary CMB signal. In this Section we investigate template fitting
and show that this provides an acceptable method for $B$-mode analysis
for \planck.

\subsection{Summary}

Let us model the data vector as
\begin{equation}
{\bf x} = {\bf s}+ {\bf F}  \pmb{$\beta$}  + {\bf n}, \label{TF1}
\end{equation}
where ${\bf s}$ is the signal, ${\bf F}$ is a template matrix,
\pmb{$\beta$} \  is a vector of unknown parameters and ${\bf n}$ is the
pixel noise vector. For example, the data vector ${\bf x}$ could be a
vector of length $2N_p$ consisting of the $Q$ and $U$ maps ${\bf x}
\equiv ({\bf Q}, {\bf U})$, \pmb{$\beta$} \ could be a vector of four
unknown amplitudes $(\beta^1_Q, \beta^1_U, \beta^2_Q, \beta^2_U)^T$, and ${\bf F}$ a $2N_p \times 4$
matrix consisting of two $Q$ and two $U$ foreground template maps
\begin{equation}
 {\bf F}  
 =  {\left ( \begin{array}{cccc}
        F^1_Q(1) & 0 & F^2_Q(1) & 0 \\
         . & 0  & . & 0 \\
         . & 0  & . & 0\\
        F^1_Q(N_p) & 0 & F^2_Q(N_p) & 0 \\
         0 & F^1_U(1) & 0 & F^2_U(1) \\
         0 & . & 0 & . \\
         0 & . & 0 & . \\
         0 & F^1_U(N_p)& 0 & F^2_U(N_p) 
       \end{array} \right ) }.  \label{TF2}
\end{equation}
We find $\pmb{$\beta$}$ by minimising 
\begin{equation}
 \chi^2 = ({\bf x}- {\bf F} \pmb{$\beta$})^T {\bf C}^{-1} ({\bf x} - {\bf F} \pmb{$\beta$}), \label{TF3}
\end{equation}
where ${\bf C}$ is the covariance matrix (\ref{like2}). The solution is
\begin{equation}
 \pmb{$\beta$}  = ({\bf F}^T{\bf C}^{-1} {\bf F})^{-1}( {\bf F}^T {\bf C}^{-1}  {\bf x}). \label{TF3a}
\end{equation}
The minimum variance estimate of the signal vector, $\bf {\hat s}$, is the Wiener-filtered
\begin{equation}
 {\bf \hat s} = {\bf S}{\bf C}^{-1} ( {\bf x} - {\bf F} \pmb{$\beta$}),  \label{TF4}
\end{equation}
(see {\it e.g.} Rybicki and Press 1992).
If the data vector is noise-free and contains zero foreground, we see that
template matching recovers a `biased' estimate of the signal,
\begin{equation}
  \hat {\bf s} =  {\bf s} - {\bf F} ({\bf F}^T {\bf C}^{-1} {\bf F})^{-1} {\bf F}^T {\bf C}^{-1} {\bf s}.  \label{TF5}
\end{equation}
This is the analogue of (\ref{ilc10}) for template matching (and is
identical for a single foreground/template if the covariance matrix
${\bf C}$ is diagonal). Notice that as with equation (\ref{ilc10}) the
offset is independent of the amplitude of the foreground
template. Even if there is no foreground in our signal, template
matching will produce a cross-correlation offset in the recovered
signal that is independent of the amplitude of the foreground.  As
with the ILC method, the cross-correlation offset gives a fundamental
irreducible limit on the amplitude of a $B$-mode component detectable by
template matching.  The critical difference with the ILC method is
that the amplitude of the offset depends on the mismatch between the
foreground matrices, $F_{Q}(i)F_{Q}(j)$ {\it etc}., and the appropriate
sub-matrices $C_{QQ}$ {\it etc}.\ of ${\bf C}$. The bigger the mismatch, 
the smaller the cross-correlation offset. The method is therefore `semi-blind'
because it uses prior information on the signal+noise covariance matrix to
determine the vector \pmb{$\beta$}. As we will show below, this prior
information reduces the cross-correlation offset by more than an order of
magnitude compared to the ILC method.

If the data vector is noise-free, but contains a foreground component
${\bf F}^\prime$, ${\bf F}^\prime \ne {\bf F}\pmb{$\beta$}$ the
minimum variance signal estimate will contain a foreground-mismatch
term in addition to the cross-correlation offset
\begin{equation}
 \hat {\bf s} =  {\bf s} - {\bf F} ({\bf F}^T 
{\bf C}^{-1} {\bf F})^{-1} {\bf F}^T {\bf C}^{-1} {\bf s} + 
[{\bf F^\prime} - {\bf F}({\bf F}^T {\bf C}^{-1} {\bf F})^{-1}
({\bf F}^T{\bf C}^{-1} {\bf F}^\prime) ].   \label{TF6}
\end{equation}
Evidently, both terms must be shown to be small for B-mode detection.  

As an illustration of the method, we have applied it to the noise-free
maps illustrated in Figure 2. We first construct  inverse noise variance
weighted $Q$ and $U$ maps (equation \ref{ilcn2}) of the CMB ($r=0$) +
PSM for the four frequency channels $70$, $100$, $143$ and $217$ GHz
using the SPP05 detector sensitivies. (The motivation for this will be
made clear in the next Section). The resulting data vector {\bf x},
although noise-free, is contaminated with both synchrotron and dust
polarization foregrounds. We then use the $30$ GHz PSM to define a
low-frequency foreground template and we use either the $217$ GHz or
$353$ GHz PSM to define a high-frequency foreground template. We then
compute the template subtracted data vector, ${\bf x} - {\bf F}
\pmb{$\beta$}$, and feed this into the likelihood function
(\ref{like1}) to compute the posterior distribution of $r$. The
results are shown in Figure 7. The $30$ GHz and $217$ GHz PSM templates
remove foregrounds to extremely high accuracy of $r \simlt 
10^{-3}$. The cross-correlation offset term in equation (\ref{TF6}) is
negligible. The primary source of error is caused by the foreground
mismatch in equation (\ref{TF6}), though for these choices of template,
the error is small. Using the $30$ GHz and $353$ GHz templates, the peak of
the likelihood is  offset by $\sim 3 \times 10^{-3}$ as a result of
foreground mismatch term at high frequencies. However, the offset
is much smaller than the width of the likelihood distribution expected
for \plancks sensitivities and is therefore ignorable. As we will see
in Section 4.3, because of  \planck's high detector noise levels, 
it is  better to use the $353$ GHz channel as a high-frequency polarization
template because it has a higher foreground signal-to-noise than the
$217$ GHz channel.

\begin{figure*}
\vskip 3.0
 truein

\includegraphics{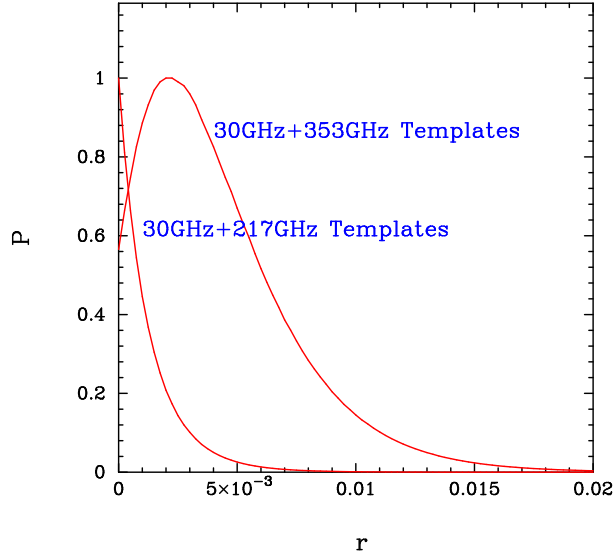}

\caption { Distributions of the tensor-scalar ratio for template
  subtracted noise-free combined frequency $Q$ and $U$ maps (frequency
  range $70$--$217$ GHz), constructed as discussed in the text. Two
  curves are shown, one using the $30$ and $217$ GHz PSM maps as
  templates and the other using the $30$ and $353$ GHz PSM.}

\label{figtemp}
\end{figure*}

\subsection{Parameter estimation with template marginalization}

In this subsection, we will discuss the problem of parameter
estimation from multiple noisy maps with foreground subtraction. The
aim is to use the results of this subsection to construct an
approximate, but easily calculable estimate of the likelihood
function.

The data ($I$, $Q$ and $U$) per pixel $i$ at each frequency is
arranged into a single column vector ${\bf X}$.  Assume the instrument
noise is described by a corresponding `large' covariance matrix ${\bf
  N}$.  The foregrounds are modelled by a a mean map at each frequency
with some spread.  We denote the mean by the vector ${\bf F_0}$ and
encode the spread via a Gaussian distribution with a covariance matrix
${\bf P}$.  (${\bf P}$ can be very general, being a `large' matrix,
but it can encode `simple' uncertainties such as a global spectral
index uncertainty, or an uncertainty about whether a particular pixel
is heavily contaminated or not.)

We assume that CMB has a blackbody, spectrum, so that the CMB-induced
signal from each pixel is given by ${\bf e} {\bf s} (i)$, or, for the
entire data-set, by ${\bf E} {\bf s}$, where ${\bf s}$ is the entire
CMB signal, and ${\bf E} \equiv {\bf I} \otimes \bf{e}$. The matrix
${\bf E}$ therefore duplicates the CMB signal ${\bf s}$ into each of
the frequency bands.  Finally, we assume that the foregrounds and CMB
are statistically independent.

Since  the CMB is assumed to be  Gaussian, the CMB signal ${\bf s}$ is
distributed as
\begin{equation}
 {1 \over \sqrt{  \vert {\bf S } \vert}} {\rm e}^{ 
-{1 \over 2}{\bf s}^T {\bf S}^{-1} {\bf s}  } \;d{\bf s}  \label{SG1}
\end{equation}
where ${\bf S}$ is the signal covariance matrix.  The model is assumed
to affect the CMB only through ${\bf S}$, so the probability for a
model may be equated to the probability for ${\bf S}$.
From our assumptions above, the prior on the foreground signal is
taken to be
\begin{equation}
  {1 \over \sqrt{  \vert {\bf P} \vert}}
{\rm e}^{ - {1 \over 2}({\bf F}- {\bf F_0})^T  {\bf P}^{-1}
    ({\bf F} -{\bf F_0})  } \;d{\bf F}.    \label{SG2}
\end{equation}
(Note we assume that ${\bf P}^{-1}$ exists, though the derivation
given below can easily be modified if it does not.)  Noise introduces
a mismatch between ${\bf X}$ and ${\bf E s} + {\bf F}$, so 
the probability density for the data given the signal and foreground
is,
\begin{equation}
 {1 \over  \sqrt{  \vert {\bf N } \vert}}
{\rm e}^{-{1 \over 2}({\bf X}- {\bf E}{\bf s}  - {\bf F})^T {\bf N}^{-1}
({\bf X}-{\bf E} {\bf s} - {\bf F})}\; {\bf dX} .    \label{SG3}
\end{equation}
To obtain the probability for a model given the data, we need to
combine equations (\ref{SG1})-(\ref{SG3}), multiply by a
prior $p({\bf S})$  and integrate over the
signal and uncertainties in the foreground model, 
\begin{equation}
p({\bf S} \vert {\bf X}) \propto
p({\bf S}) 
\int\int d{\bf F} d{\bf s}
 {1 \over  \sqrt{ \vert {\bf N } \vert}}
{\rm e}^{-{1 \over 2}({\bf X}- {\bf E}{\bf s}  - {\bf F})^T {\bf N}^{-1}
({\bf X}-{\bf E} {\bf s} - {\bf F})}
{1 \over \sqrt{\vert {\bf P} \vert}} 
{\rm e}^{ - {1 \over 2}({\bf F}- {\bf F_0})^T  {\bf P}^{-1}
    ({\bf F} -{\bf F_0})  }
{1 \over \sqrt{ \vert {\bf S } \vert}} {\rm e}^{ 
-{1 \over 2}{\bf s}^T {\bf S}^{-1} {\bf s}  } . \label{SG4}
\end{equation} 
From now on we will drop ${\bf S}$-independent factors without comment.  Writing
the exponent as 
\begin{equation}
 -\frac{1}{2} ({\bf s}^T \;\; {\bf F}^T) 
  {\left ( \begin{array}{cc}
        {\bf E^T}{\bf N}^{-1} {\bf E}+ {\bf S}^{-1} &  {\bf E}^{T}{\bf N}^{-1} \\
         {\bf N}^{-1} {\bf E} & {\bf N}^{-1} + {\bf P}^{-1} 
       \end{array} \right ) } 
{ \left( \begin{array}{c} {\bf s} \\ {\bf F} \end{array}
\right ) }  +   ({\bf s}^T \;\; {\bf F}^T) 
 {\left ( \begin{array}{c}
        {\bf E^T}{\bf N}^{-1} {\bf x} \\
         {\bf N}^{-1} {\bf x} + {\bf P}^{-1}{\bf F_0}^{-1} 
       \end{array} \right ) },  \label{SG5}
 \end{equation}
performing the integral over ${\bf s}$  and ${\bf F}$ and using 
the formulae for block matrices and determinants in Press \etals (1992)
the result simplifies to the compact expression
\begin{equation}
{p({\bf S }) \over \sqrt{ \vert {\bf S} + {\bf M} \vert}} {\rm  e}^{-\frac{1}{2} {\bf Y}^T ( {\bf S}
+ {\bf M})^{-1} {\bf Y}},
\label{SG6}
\end{equation}
where we have defined ${\bf M}$ via
\begin{equation}
{\bf M}^{-1} \equiv {\bf E^T} ( {\bf N}+ {\bf P})^{-1} {\bf E}\label{SG7}
\end{equation}
and  ${\bf Y}$ is
\begin{equation}
{\bf Y} \equiv {\bf M} {\bf E}^T ({\bf N}+ {\bf P })^{-1} ( {\bf x} - {\bf F_0}).
\label{SG8}
\end{equation}
Note that equation (\ref{SG7}) has a simple interpretation. It is just
the usual likelihood function (\ref{like1}) applied to a map ${\bf Y}$
constructed from the data with a noise covariance matrix ${\bf M}$
give by equation (\ref{SG7}). Furthermore, the map (\ref{SG8}) is
basically an inverse-noise weighted map over the frequency channels
(but note the `noise' here is the sum of the detector noise ${\bf N}$
and the foreground uncertainty ${\bf P}$ and that the best-guess foreground ${\bf F_0}$ is
subtracted from  the data at the start).

This derivation provides the justification for the following simplified
model for a polarization likelihood for \planck. In Section 4.1, we
demonstrated that by using two templates, foreground-mismatch could be
reduced to levels that are negligible compared to the \plancks
instrumental noise. It is therefore a good approximation to neglect
${\bf P}$ in comparison to ${\bf N}$. The map ${\bf Y}$
can then be approximated  by subtracting the best fit templates (as 
computed in Section 4.1) from an inverse noise-variance weighted
set of maps. This justifies the procedure used to construct Figure
7.

Gratton (2008) presents a conceptually similar technique for
constructing a likelihood function from noisy data in the presence of
foregrounds.  Given assumed priors, the scheme effectively averages
over linearly weighted combinations of frequency channels.  The
advantage of that technique is that it self-consistently marginalizes
over foregrounds, rather than selecting specific channels as templates
as described above.  In fact, in the noise-free limit, to the extent
that (\ref{TF1}) (with ${\bf n}=0$) is an accurate model of the data,
the technique gives best-fit weights, $w_i$, that also yield (\ref{TF5}) as
the most likely CMB map, mitigating foreground mismatch.
However, with noise, the algorithm for the resulting likelihood
function is considerably more complicated.  A detailed comparison
between the two approaches is currently in progress.

\subsection{Application to simulations with noise}

In this subsection, we apply the above formalism to simulations that
include the \plancks noise levels. The detector noise is assumed to be
white and uncorrelated in each pixel. It would be straightforward to
generalise this analysis to include a more realistic scanning strategy
and correlated `destriping' noise (all that is required is an
appropriate model for the noise covariance matrices $N_{ij}$ that
includes destriping errors). However, correlated errors are expected
to be small for \plancks and so we ignore them in this analysis (for a
detailed discussion see Efstathiou 2007). We therefore add
uncorrelated white noise to the $NSIDE=2048$ primary CMB + foreground
maps at each of the polarization-sensitive \plancks frequencies. The
maps are smoothed with a Gaussian of $7^\circ$ FWHM and reconstructed
at $NSIDE = 16$. The low resolution noise covariance matrices are
computed using the small-angle approximation (see Appendix A), which
gives an excellent approximation to the true covariance matrices.

\begin{figure*}
\vskip 7.0 truein

\includegraphics{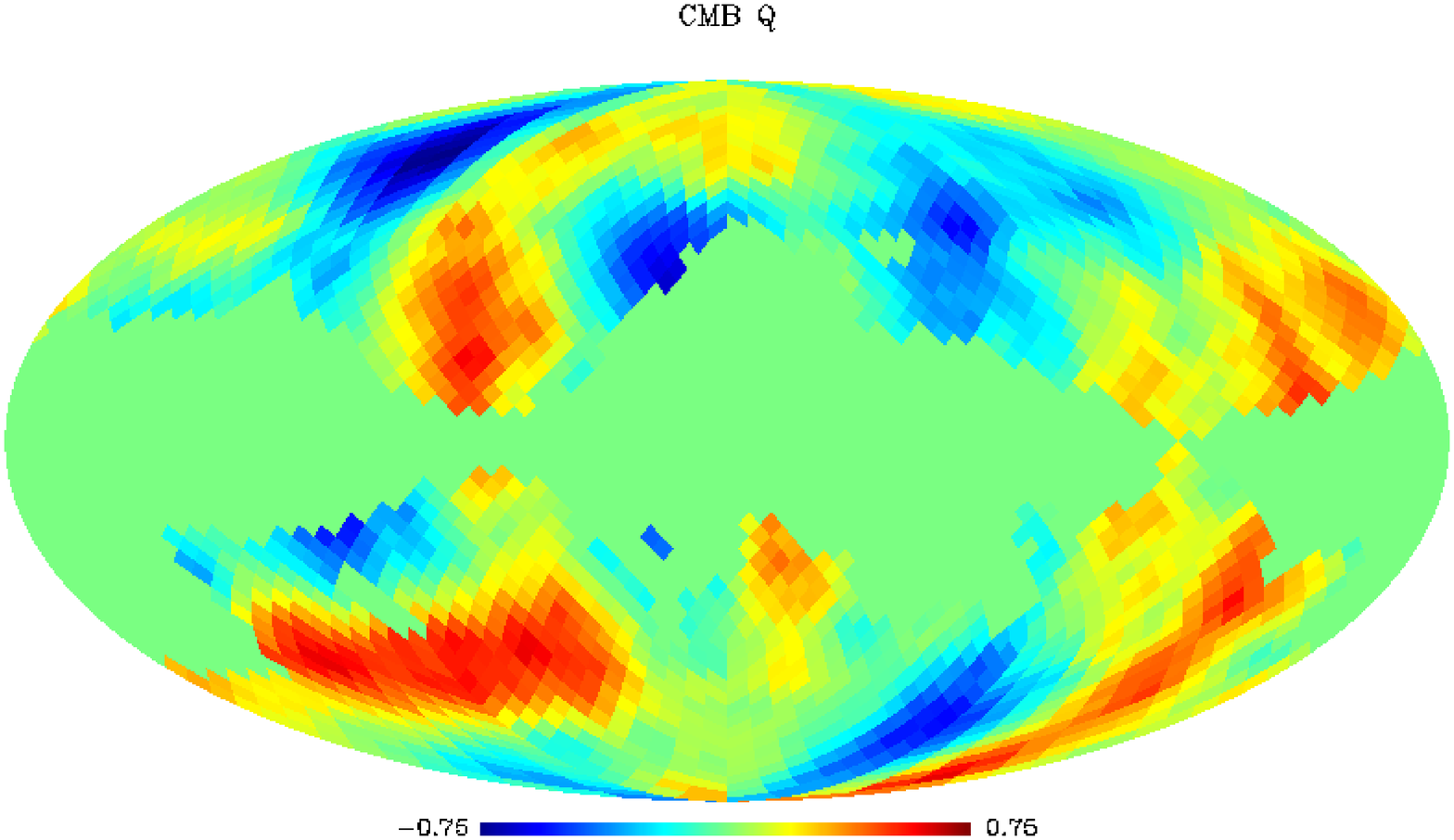}
\includegraphics{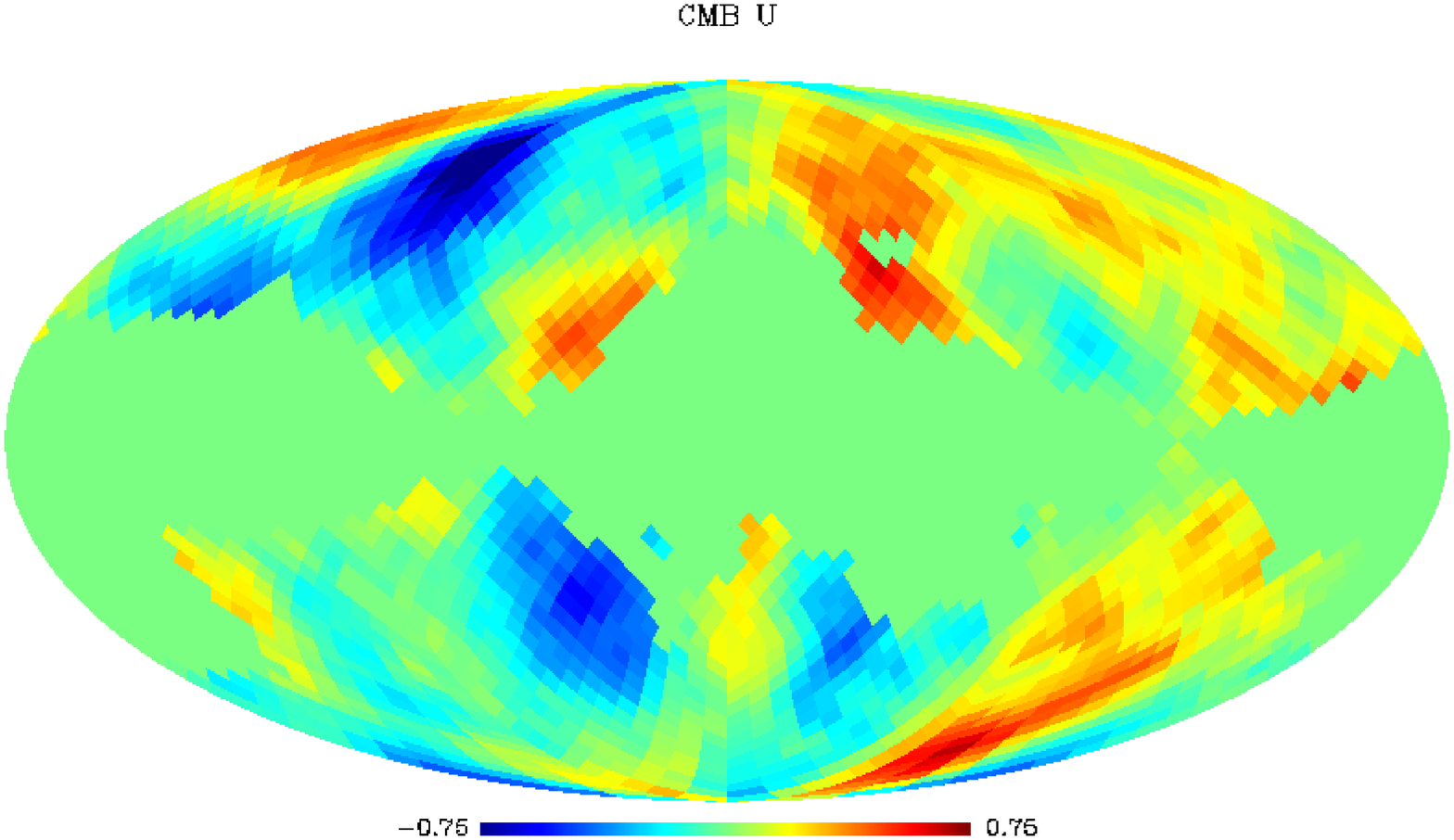}
\includegraphics{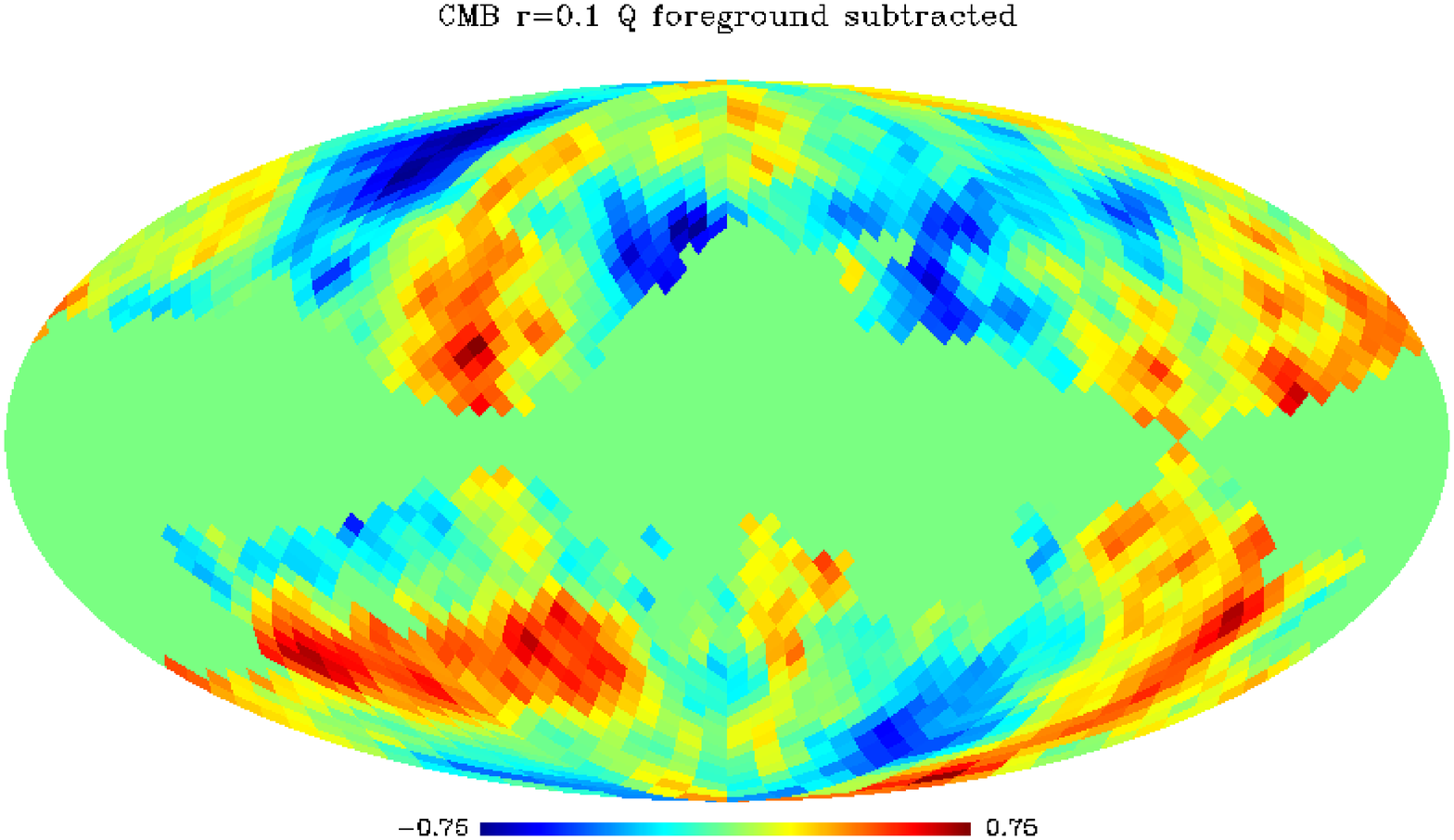}
\includegraphics{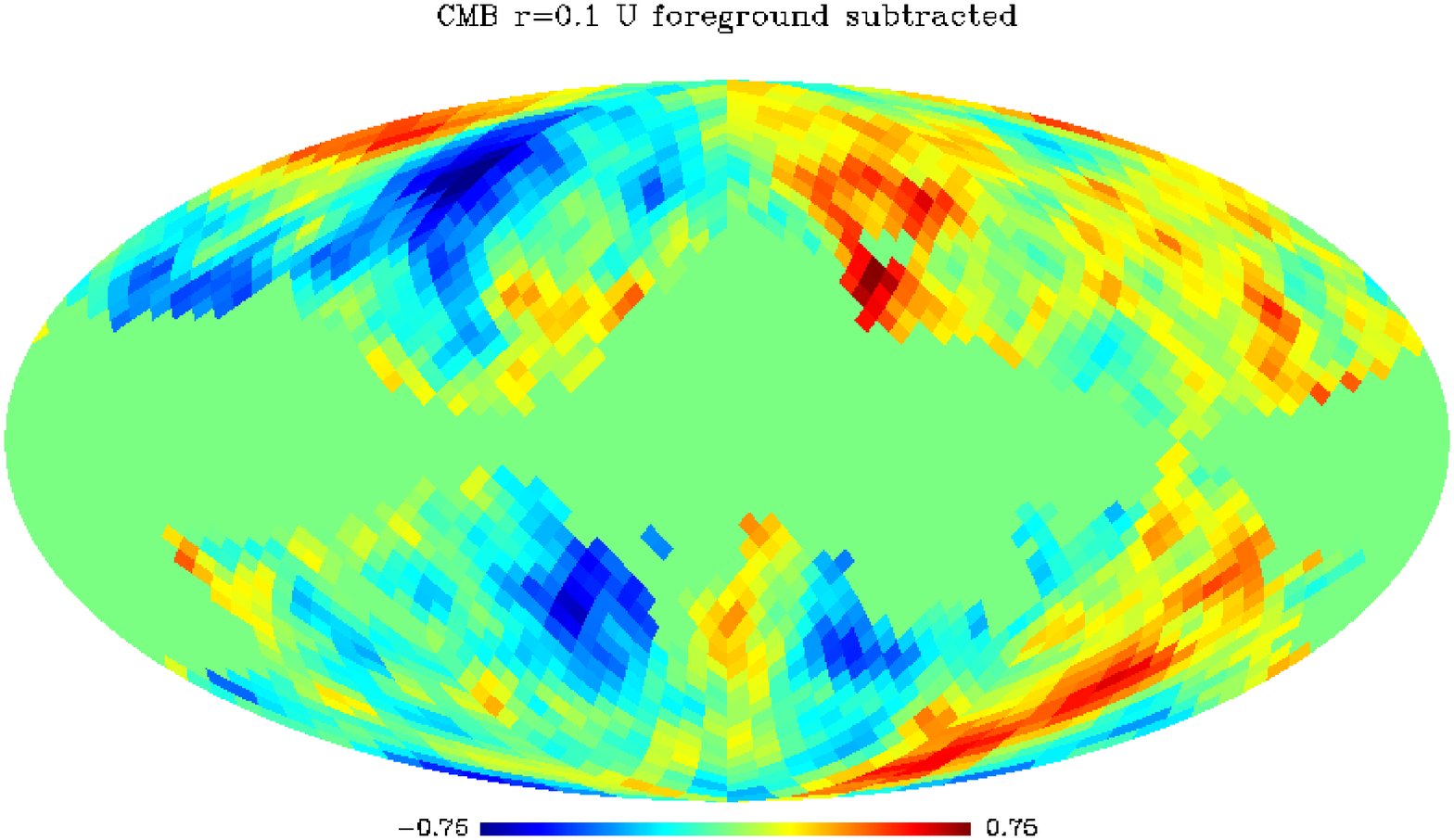}
\includegraphics{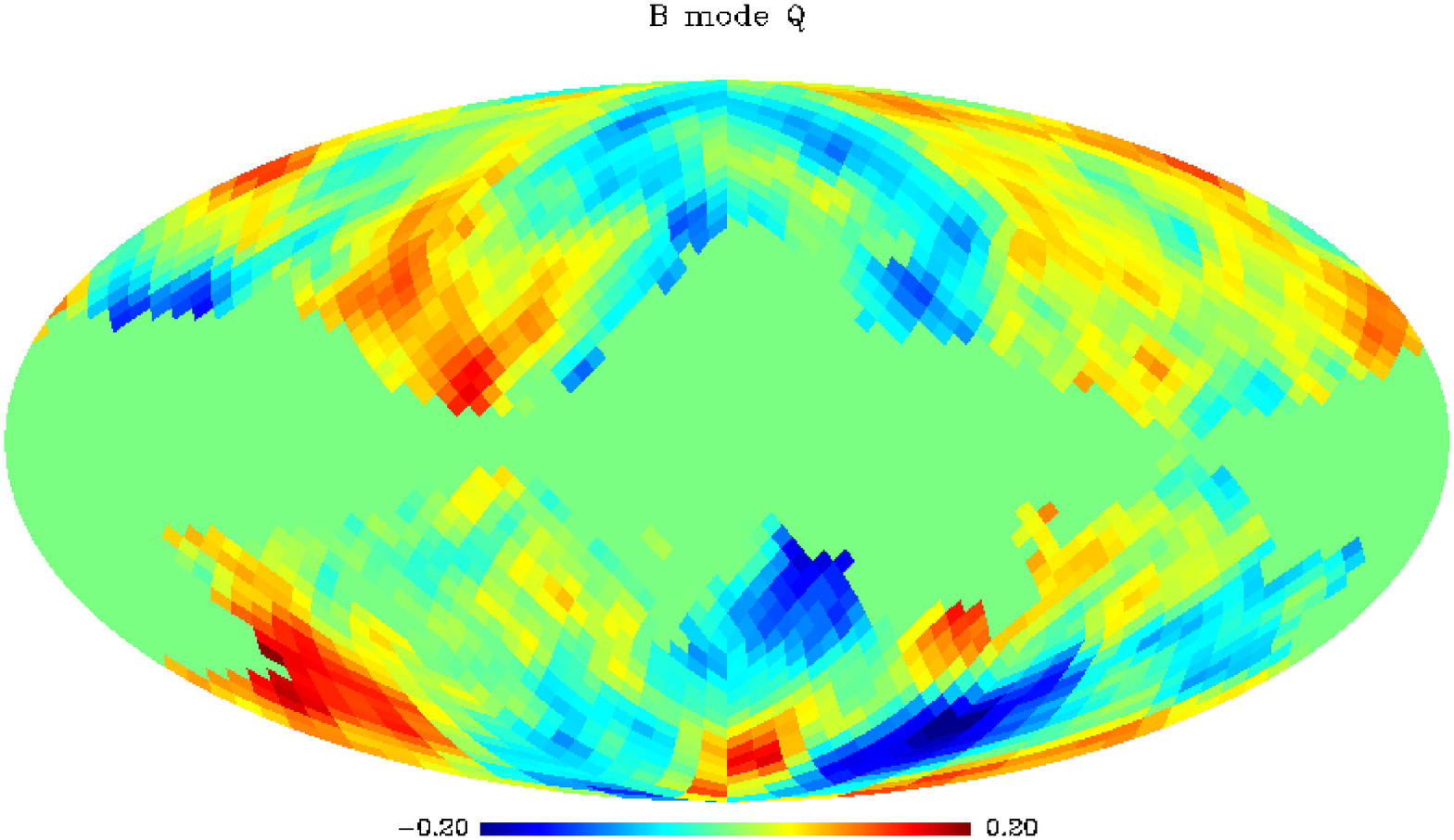}
\includegraphics{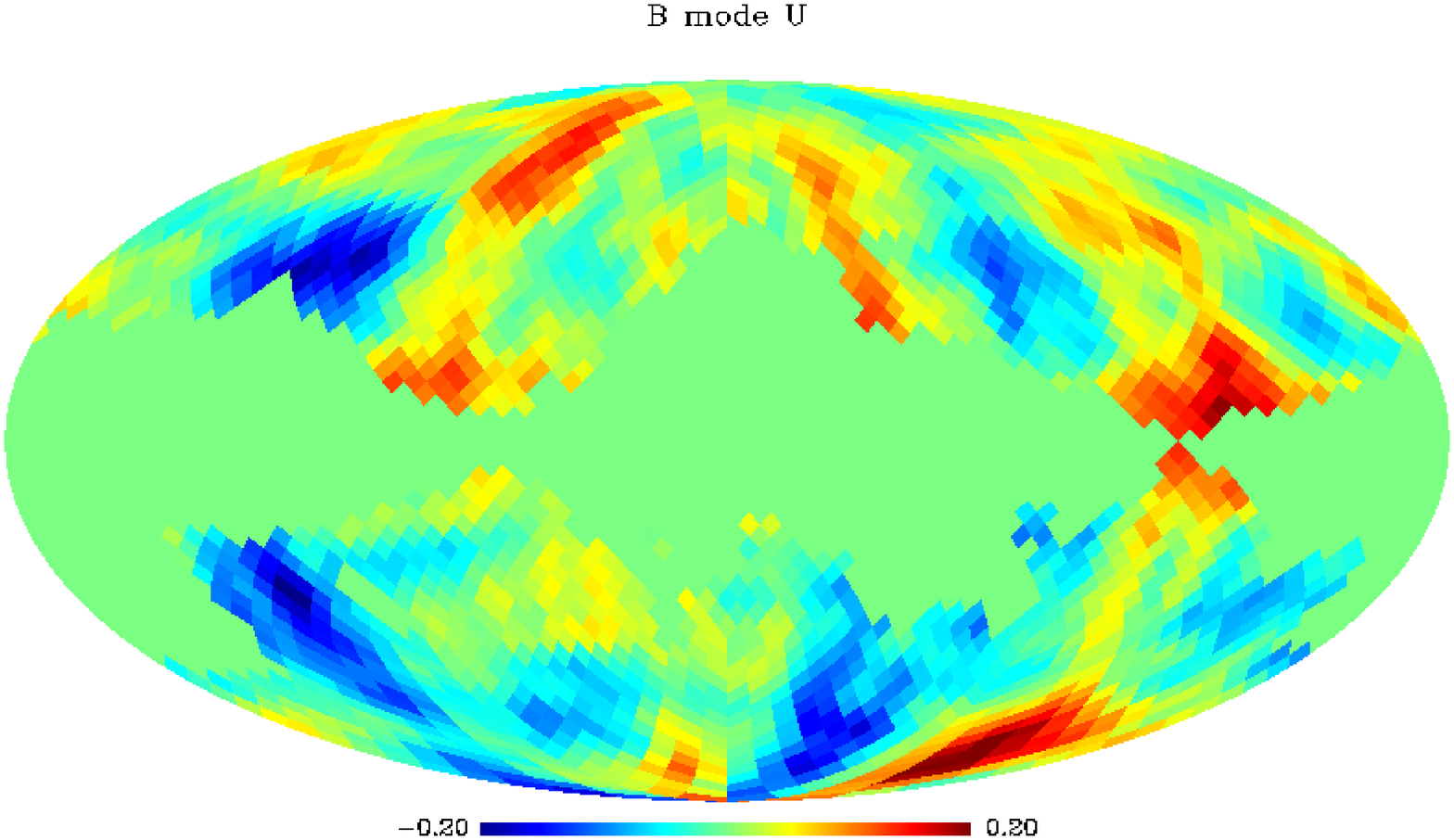}
\includegraphics{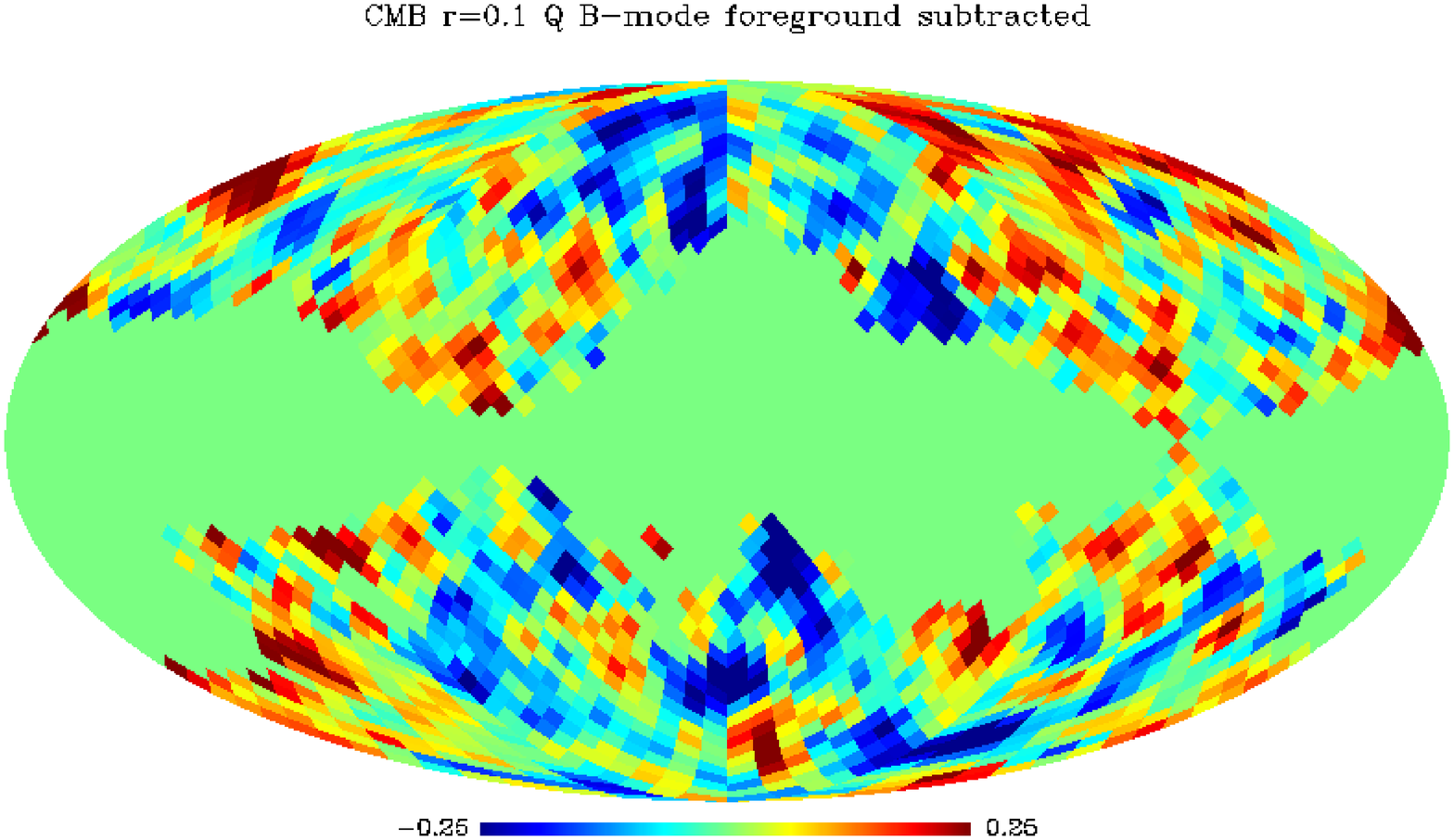}
\includegraphics{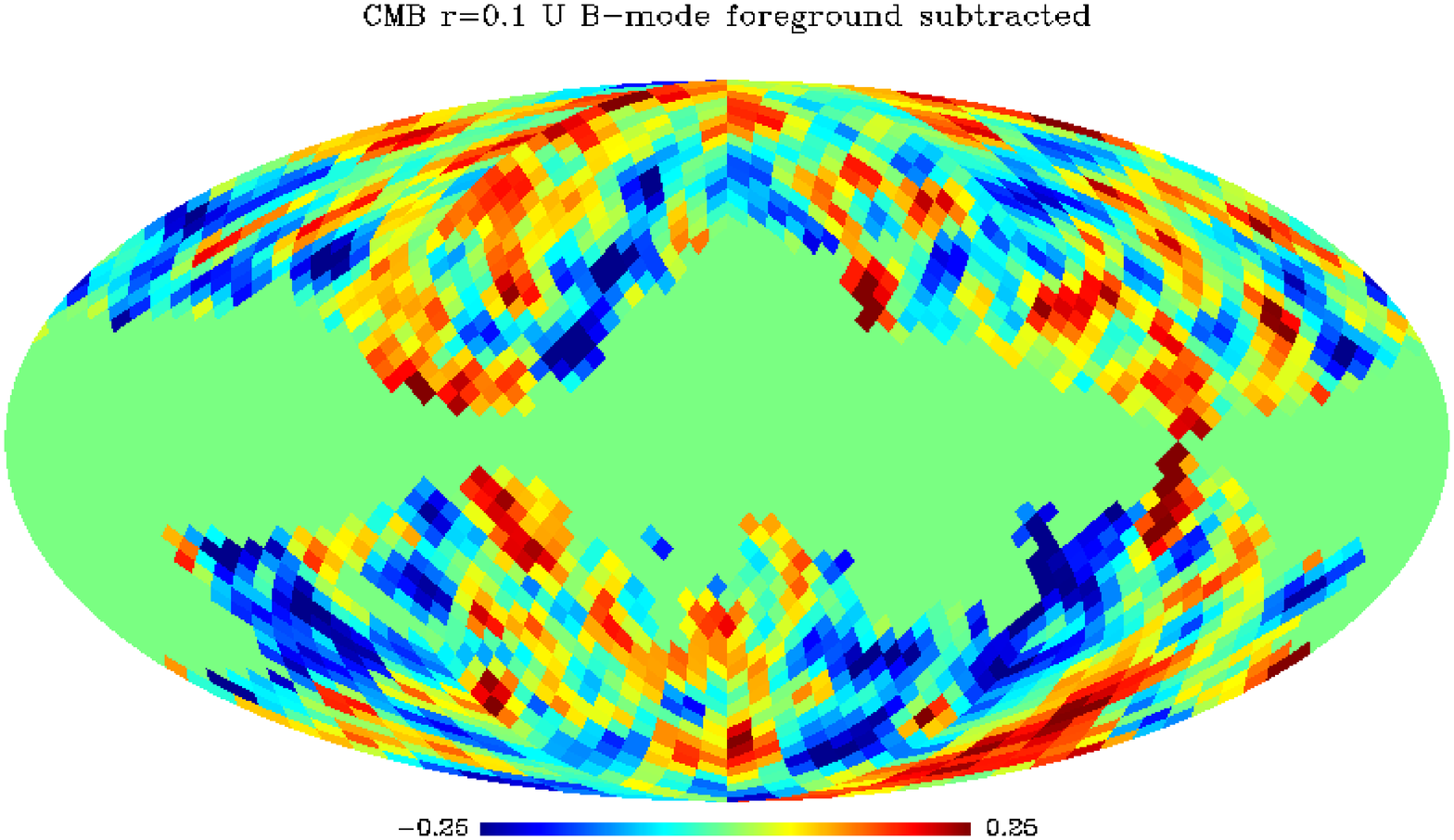}

\caption {Q and U maps: Upper panel shows the noise-free CMB
  simulations with $r=0.1$ for regions outside the internal mask.  The
  second panel shows the foreground subtracted noisy reconstructions
  computed as described in the text. The third panel shows the $B$-mode
  contribution to the noise-free CMB maps. The lowest panel shows the
  noisy foreground subtracted reconstruction of this $B$-mode
  contribution.}

\label{figure8}
\end{figure*}

In reality, the templates will contain a primary CMB
signal\footnote{It is unlikely that `external' templates would be of
  any value for $B$-mode analysis.} and so the likelihood
approximation is slightly more complicated than implied in Section
4.2. As described above, the data vector $x_i$ is constructed as
an inverse noise variance weighted sum over a set of frequency channels.
The covariance matrix of this vector is written as 
\begin{equation}
\langle x_ix_j \rangle= S_{ij}+ \Phi_{ij}+ N_{ij}, \label{TFF1}
\end{equation}
where $ S_{ij}$, $\Phi_{ij}$ and $N_{ij}$ are respectively the
primordial CMB, residual foreground and noise covariance matrices.
Now construct the data vector
\begin{equation}
Y_i  = x_i -F^k_i \beta^k_i,  \qquad  ( 
\beta^k_i = \beta^k_{(Q, U)}, {\rm if} \ i \equiv (Q, U)  ) , \label{TFF2}
\end{equation}
where the superscript denotes frequency.  If the template subtraction
removes the foregrounds, the average of (\ref{TFF2}) over
noise-realizations is
\begin{equation}
\langle Y_i \rangle = s_i(1 - \sum_k \beta^k_i) , \label{TFF3}
\end{equation}
and  if the coefficients $\pmb{$\beta$}$ \ are independent of
the signal the covariance matrix  $\langle Y_i Y_j \rangle$ is
\begin{equation}
\langle Y_iY_j \rangle = S_{ij}(1 - \sum_k \beta^k_i)(1 - \sum_k \beta^k_j) + N_{ij} + N^k_{ij}\beta^k_i\beta^k_j. \label{TFF4}
\end{equation}
The solution for $\pmb{$\beta$}$ \ is found by iteratively minimising
(\ref{TF3}) with ${\bf C}$ replaced by $\langle Y_iY_j \rangle$ and
ignoring any weak correlation between the solution and the signal. The
final data vector ${\bf Y}$ and its covariance matrix (\ref{TFF4}) are
then used to compute the likelihood function (\ref{like2}). The parameters
$\pmb{$\beta$}$ are well constrained by the data and so it is a good
approximation to keep them fixed at their central values. The main
contribution of the  $\pmb{$\beta$}$ to the error budget is via the 
noise term (\ref{TFF4}).

\begin{figure*}
\vskip 3.0
 truein

\includegraphics{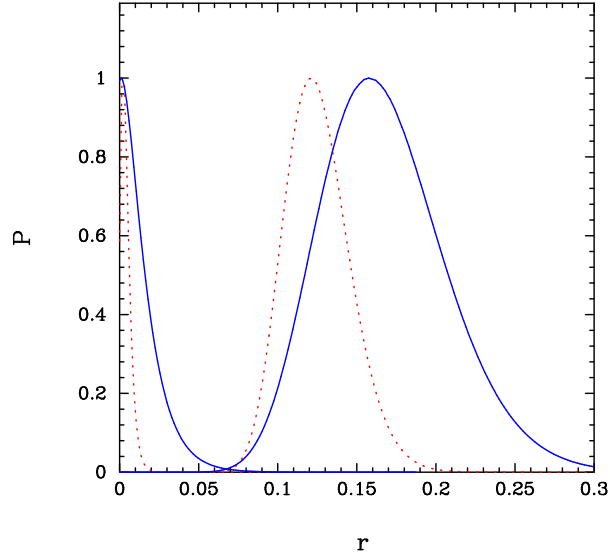}

\caption { Distributions of the tensor-scalar ratio for two simulations with
$r=0$ and $r=0.1$. The dotted (red) lines show the distributions for
noise-free and foreground-free simulations for regions outside the internal
mask. The solid (blue) lines show the distributions for foreground
subtracted noisy simulations, as described in the text.}

\label{likefinal}
\end{figure*}

In the simulations described here, we construct the data vector ${\bf
  x}$ from the four frequency channels $70$, $100$, $143$, $217$ GHz,
since there is little additional sensitivity to primary CMB signal in
the other channels. We use the $30$ GHz and $353$ GHz channels as
templates. The internal mask described in Section 2 is applied to all
channels. The resulting noisy foreground subtracted maps are shown in
Figure 8. The upper panel of this figure shows the noise-free CMB
simulations for $r=0.1$ for the regions that lie outside the internal
mask. The second panel shows the reconstruction after  foreground
subtraction from the noisy maps following the procedure described
above. There is clearly a very good correspondence between the two
sets of maps. The third panel in Figure 8 shows the noise-free
contribution of the $B$-mode to the $Q$ and $U$ maps. The foreground
subtracted reconstruction is shown in the lowest panel. Again, there
is a good correspondence between the maps, but the reconstructed maps
are very noisy. In fact, instrument noise dominates over foreground
mismatch. A substantial component of the noise comes from the
templates because the $30$ and $353$ GHz channels of \plancks are
significantly noisier than the main `CMB' channels at $\sim 100$ GHz.

The likelihood functions for $r$ are shown in Figure 9. The dotted
(red) lines show the likelihood functions applied to the noise-free
CMB maps (though with diagonal `regularizing' noise applied, as
described in Section 3) for the two simulations with $r=0$ and $r=0.1$
for the regions outside the internal mask. These likelihoods are close
to the `best' that could be achieved from a low resolution experiment
in the absence of foreground contamination.  The results from our
noisy foreground subtracted simulations are shown by the solid (blue)
lines. The distribution for the model with $r=0$ is peaked close to
$r=0$, so clearly residual foreground mismatch is unimportant. The
increased widths of the blue curves are caused by residual instrument
noise, including the noise in the template channels. All of the
results described in this paper assume a nominal mission lifetime of
$14$ months for \planck. This ensures that every detector on \plancks
covers the sky twice (see SPP05). We show in a separate paper
(Efstathiou and Gratton, 2009) that an extended mission lifetime for
\plancks to four full sky surveys leads to a significant improvement
in $B$-mode sensitivity.

Finally, Figure 10 shows the QML power spectra (equations
\ref{ML1}-\ref{ML4}) for the foreground subtracted noisy realizations
with $r=0$ and $r=0.1$. As discussed in Efstathiou (2006), the QML
estimator eliminates mixing of $E$ and $B$-modes at low multipoles
almost perfectly on a cut-sky, and so the $B$-mode spectrum for the
$r=0$ realization is indeed close to zero at low multipoles.  However,
this Figure shows clearly that for the \plancks noise levels, there is
little information in the $B$-mode spectrum at multipoles $\ell \simgt
10$.

\begin{figure*}
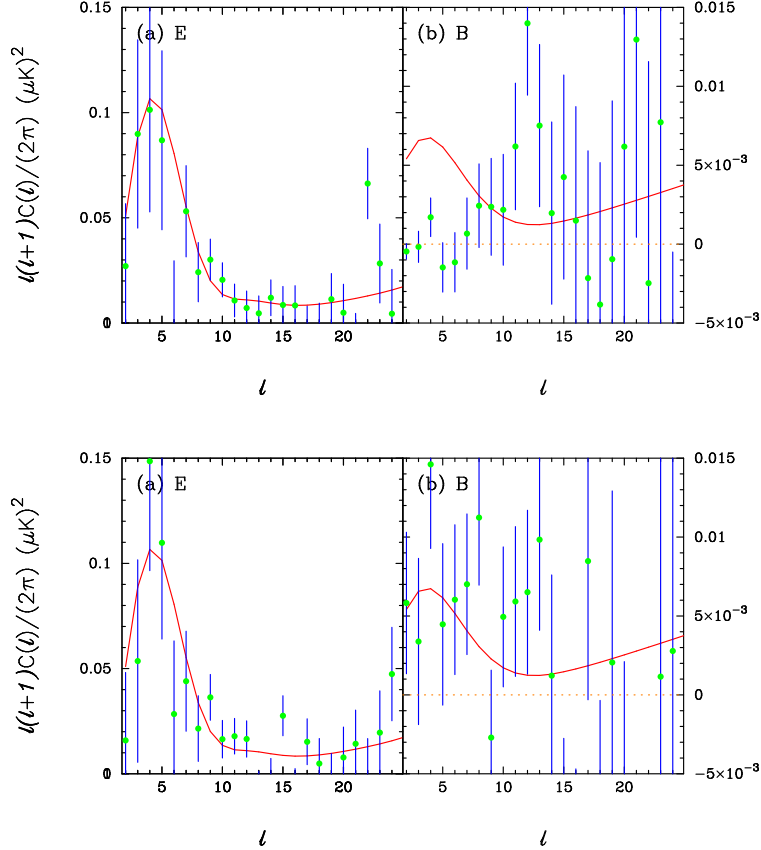

\vskip 4.5
 truein

\includegraphics{pgqml0.0.ps}

\includegraphics{pgqml0.1.ps}

\caption {QML estimates of the $E$ and $B$-mode polarization spectra
  for the two noisy foreground subtracted simulations used to compute
  the likelihoods of Figure \ref{likefinal}.  The upper panel shows
  the power spectra for the realization with $r=0$ and the lower panel
  shows the power spectra for the realization with $r=0.1$. The error
  bars show the diagonal components of (\ref{ML4}) using the
  theoretical input values of $r$ for each realization.  The solid
  lines show the theoretical input spectra for $r=0.1$.}

\label{qmlfinal}
\end{figure*}

\section{Comments on foreground removal techniques}

As summarized  in the Introduction, a large number of diffuse foreground
subtraction techniques have been discussed in the literature. Some of
these are designed to recover `physical' foregrounds, {\it e.g.}
separating free-free from synchrotron emission. Other techniques are
designed to tackle the problem described in this paper, {\it i.e.}
the recovery of the primordial CMB anisotropy from foreground
contaminated maps. The remarks in this Section apply to this latter
class of techniques.

In the preceeding Sections, we identified two distinct forms of error,
which we termed `cross-correlation offset' and `foreground mismatch'.
These two types of error provides an intuitively useful way of
classifying component separation methods. Table 3 summarizes our
proposed classification scheme. Any foreground subtraction technique
can be placed somewhere between the blind and unblind rows of this
Table.

Briefly, a purely blind technique such as ILC can, in
theory\footnote{{\it i.e.} a set of weights exists.}, reduce
foreground mismatch to negligible levels provided there are enough
frequency bands to describe the foregrounds. (This is the case, for
example, for the weights listed in the first three rows of Table
2). However, in a purely blind technique, there is no external
information to distinguish between CMB and foreground components with
similar structure on the sky. The result is a
cross-correlation offset that is independent of the amplitude of the
foregrounds. Any purely blind technique ({\it e.g.} harmonic ILC) will
show a cross-correlation offset. Unless one can isolate pure $B$-modes
on a cut sky (see {\it e.g.} Lewis 2003), cross-correlation between
the CMB $E$-modes and the foregrounds can produce a potentially 
serious cross-correlation offset.

The amplitude of the cross-correlation offset can be reduced if some
additional information is provided. We have classified template
matching as a `semi-blind' technique because it makes use of some
prior information, though it is not based on a physical model of the
foregrounds. The method requires a model for the signal (primordial
CMB) covariance matrix and the templates provide a model for the
angular distribution of the foregrounds (spectral index variations can
be taken into account by adding more templates). As shown in the
previous Section it is possible to reduce both cross-correlation offset
and foreground mismatch to negligible levels.

A third class of technique attempts to model the foregrounds by
fitting a parametric physical model (Brandt \etals 1994; Eriksen \etals
2006; Dunkley \etals 2008b). If the physical model is a correct
representation of the truth, it is possible to reduce both
the cross-correlation offset and foreground mismatch to negligible
levels. However, this type of technique is limited, in practice, by
the number of frequency channels available.  The number of independent
parameters describing the model must be less than or equal to the
number of frequency channels. For \plancks polarization, this limits
the number of independent free parameters to be $\le 7$ (if the $Q$
and $U$ models are treated independently). This limits the complexity
of the physical model, limiting the scope for redundancy checks. Of
course, if the model is incorrect the method will be limited by
foreground mismatch.

\begin{table}
\centerline{\bf \ \ \  Table 3:  Classification scheme of foreground removal techniques}

\begin{center}

\begin{tabular}{ccc} \hline \hline
\smallskip 
Scheme       & cross-correlation offset & foreground mismatch \cr
\hline \hline
Blind (e.g. ILC)  & Significant & Small (given enough frequency bands)    \cr
Semi-blind (e.g. template fitting)  & Small  &  Small (given enough templates) \cr
Unblind (e.g. model fitting)     & Small (if model is correct) &  Small (if model is correct) \cr
\hline
\end{tabular}

\end{center}
\end{table}

It is also useful to consider how foreground subtraction techniques
are affected by instrumental noise. Instrumental noise in a purely
blind technique is, in a sense, `uncontrollable'. For example, the
`ideal' weights listed in the first three rows of Table 2 remove
foregrounds to high precision. However, if they were applied to noisy
data they would amplify the instrumental noise to high levels (because
many of the weights exceed unity). For \plancks polarization, the
resulting noise amplification would be unacceptable.  We have shown in
Section 4.2 that in the template matching approach, instrumental noise
is `controllable' provided the templates have high
signal-to-noise. Instrumental noise is a major problem for model
fitting techniques. As far as we are aware, nobody has yet developed a
model fitting technique that incorporates prior information on the
angular variation of the spectral indices of the diffuse foregrounds
(which vary slowly over the sky). As a proxy, model fitting is usually
done by independently fitting parameters in a very coarsely pixelized
map. This reduces the effects of instrument noise on the estimated
parameters, but even then the effects of noise can be limiting. For
example, in the analysis of the WMAP 5-year polarization data at a
resolution of $NSIDE=8$ the synchrotron spectral index was computed at
a resolution of $NSIDE=2$ ($48$ pixels over the whole sky) in Dunkley
\etals (2008a).  Fairly strong (though not unreasonable) priors were
imposed to find convergent solutions ({\it e.g.} the dust spectral
index, which is poorly constrained by the data, was kept fixed).
Nevertheless, Dunkley \etal's results for the $E$-mode power spectrum
at low multipoles compare well with those from the template cleaned
maps of Gold \etals (2008). It remains to be seen whether model
fitting can perform well for the more difficult problem of $B$-mode
recovery for \planck. We hope to report on this in a future paper.

\section{Conclusions}

In this paper, we have used the Planck Sky Model to assess the impact
of foregrounds on B-mode detection by \plancks at low multipoles. We
have analyzed the internal linear combination technique and shown that
the offset caused by $E$-mode polarization pattern (cross-correlation
offset) leads to a fundamental limit of $r \sim 0.1$ for the
tensor-scalar ratio even in the absence of instrumental noise. This is
comparable to the sensitivity limit of \plancks if foregrounds are
neglected. For realistic \plancks instrument noise, ILC amplifies the
noise of the `cleaned' polarisation maps to unacceptably high levels.
Our results show that `blind' techniques such as ILC are unsuitable
for detecting primordial $B$-modes from a future low-noise `CMBpol'
mission.

We have analysed template fitting, using internal templates
constructed from the \plancks data and devised a scheme to approximate
the likelihood function (\ref{like1}) from multi-frequency maps. We
have shown that this scheme works well for \plancks and offers a
feasible way of recovering primordial $B$-modes from dominant
foreground contamination even in the presence of noise. According to
the results shown in Figure 9, \planck, after the nominal mission
lifetime of $14$ months, could set a useful upper limit of $r \simlt
0.05$ if there is no primordial tensor mode and may even detect a
tensor mode if $r \sim 0.1$.  This provides a useful complement to
ground based/sub-orbital experiments which cannot probe these low
multipoles ($\ell \simlt 10$). These limits probe an interesting part
of parameter space (see Efstathiou and Chongchitnan, 2006, for a
review). For  inflation with a power law potential, $\phi^\alpha$,
the scalar spectral index and tensor-scalar ratio are approximately
\begin{equation}
  n_s \approx 1 - {2 +\alpha \over 2N}, \qquad  r \approx { 4 \alpha \over N},  \qquad {\it i.e.} \;\; r \approx 8(1- n_s) {\alpha \over \alpha + 2}, \label{C1}
\end{equation}
where $N$ is the number of inflationary e-folds between the time that
CMB scales crossed the `horizon' and the time that inflation
ends. There are indications from WMAP and CMB experiments probing
higher multipoles for a small tilt\footnote{The significance of this
  tilt depends quite sensitively on the complexity of the model (and
  assumed priors), for example on whether or not a tensor mode or a run in the
  spectral index are included in the model.} of the spectral index
$n_s \sim 0.97$ (Komatsu \etals 2008; Reichart \etals 2008). If this
tilt is correct, then the last of these equations suggests $r
\sim 0.1$ for any $\alpha$ of order unity. For example, for $N \approx
60$ (Liddle and Leach 2003), the quadratic potential (still allowed by
the data) gives $n_s \approx 0.97$, $r \approx 0.13$, within the
parameter range accessible to \planck. Failure to detect a $B$-mode at
$r \simgt 10^{-2}$ would put pressure on `high-field' ($\phi$ of order
Planck scale) inflation models, in which there has been recent renewed
interest (Silverstein and Westphal 2008).

The simplicity of the PSM is a source of concern. Following WMAP,
there is quite a lot of information available on the polarized
synchrotron emission.  The low frequency channels on \plancks will
provide additional information at $\simlt 70$ GHz. Thus, there is
considerable scope for redundancy checks at low frequencies, for
example, by varying templates and by parameter fitting. The dust
contribution to polarization is much more uncertain. Neither the level
of polarization, nor dust spectral index variations are well
constrained by current data.  We will almost certainly have to wait until
\plancks flies to assess whether polarized dust emission poses a
serious problem for $B$-mode analysis.  \plancks is heavily reliant on
the $353$ GHz channel to model dust polarization, because it has the
highest signal-to-noise on the foreground. There is some limited scope
for redundancy checks using the $217$ GHz channel. However, if
polarized dust emission is complex, it may not be possible to achieve
a limit of $r \sim 0.1$ with \plancks using template fitting or any
other foreground removal technique.

\vskip 0.1 truein

\noindent {\bf Acknowledgments:} GPE and SG thank STFC for financial
report.  The authors acknowledge the use of the Planck Sky Model
developed by the Component Separation Working group of the Planck
Collaboration, and of the Healpix package. We thank Anthony Challinor,
Jacques Delabrouille, Jo Dunkley, Antony Lewis, Hiranya Peiris and the
BICEP collaboration for useful discussions.

\begin{appendix}

\section{Pixel noise covariances of degraded resolution maps in the small angle limit}

\subsection{Temperature}

Let $x_i$ denote the pixel value in the high resolution map and $X_i$ denote the
pixel value in the low resolution map. The harmonic coefficients computed from
the high resolution map is
\begin{equation}
  a_{\ell m} = \sum_i x_i \Omega_i Y^*_{\ell m} ({\bf \theta}_i),  \label{App1}
\end{equation}
where $\Omega_i$ is the solid angle of a high resolution map pixel.
So, the pixel values in the degraded map are 
\begin{equation}
  X_i = \sum_{\ell m} a_{\ell m} Y_{\ell m}({\bf \theta}_i) f_\ell (\theta_s),  
\label{App2}
\end{equation}
where $f_\ell(\theta_s)$ is the smoothing function applied to the high resolution
map. In terms of the pixel values of the high resolution map,
\begin{equation}
  X_i = \sum_{\ell m p} x_p \Omega_p Y^*_{\ell m}({\bf \theta}_p)  Y_{\ell m} (
{\bf \theta}_i) f_\ell (\theta_s).
\label{App3}
\end{equation}
Using the addition theorem for spherical harmonics, the pixel noise covariance of
the low resolution map is
\begin{equation}
\langle X_i X_j \rangle = \sum_{\ell_1 \ell_2} \langle x_p x_q \rangle
{(2 \ell_1 + 1) \over 4 \pi}{(2 \ell_2 + 1) \over 4 \pi} \Omega_p \Omega_q
P_{\ell_1} (\cos \theta_{ip})  P_{\ell_2} (\cos \theta_{jq}) f_{\ell_1} f_{\ell_2}.
\label{App4} 
\end{equation}
This expression is time consuming to evaluate, but it simplifies significantly if
the noise is diagonal $\langle x_p x_q \rangle = \sigma^2_p \delta_{pq}$ and if we
assume small angles. In this case, for a Gaussian smoothing function, equation
(\ref{App4}) simplifies to 
\begin{equation}
\langle X_i X_j \rangle \approx {1 \over \theta_s^4} {1 \over 2 \pi^2}
\sum_p \sigma^2_p \Omega^2_p {\rm exp} \left ( - {\theta^2_{ip} \over 2 \theta^2_s} \right ){\rm exp} \left ( - {\theta^2_{jp} \over 2 \theta^2_s} \right). \label{App5}
\end{equation}

\subsection{Polarization}

In the case of diagonal pixel noise at high resolution,  and for small angles, the
 polarization covariance matrices $\langle Q_i Q_p \rangle$ for the degraded
resolution maps can be approximated by equation (\ref{App5}). It is interesting to
see why this is so. We will consider degraded resolution
$Q$ maps (the analysis is identical for $U$ maps). The equivalent to (\ref{App3})
in obvious notation is
\begin{equation}
Q_i = {1 \over 2} \sum_{\ell m} \sum_p (q_p \left [ \;_2Y^*(p)\;_2Y(i)
+ \;_{-2}Y^*(p) \;_{-2}Y(i) \right] + i u_p \left [ \;_2Y^*(p)\;_2Y(i)
- \;_{-2}Y^*(p) \;_{-2}Y(i) \right]) \Omega_p f_\ell. \label{App6}
\end{equation}
Now the  addition theorem for the tensorial harmonics is
\begin{equation}
\sum_m \;_{s_1} Y^*_{\ell - m} (\theta_1, \phi_1) \;_{s_2} Y_{\ell -m}
(\theta_2, \phi_2) = (-1)^{s_1}  \left ( {2 \ell + 1 \over 4 \pi} \right)^{1/2}
\;_{-s_1} Y_{\ell s_2} (\beta, \alpha) {\rm e}^{i s_1 \gamma}, \label{App7}
\end{equation}
where we use the Euler angle conventions of Varshalovich, 
Moskalev and Khersonskii (1988). Applying the addition theorem to
(\ref{App6}) gives
\begin{equation}
Q_i = \sum_{\ell p} {1 \over 2}  \left ( {2 \ell + 1 \over 4 \pi} \right)^{1/2}
\left (q_p \left [ \;_{-2}Y_{\ell 2} {\rm e}^{i 2 \gamma} + \;_{2}Y_{\ell -2} { \rm e}^{-i2 \gamma} \right] +iu_p\left [ \;_{-2}Y_{\ell 2} {\rm e}^{i 2 \gamma} - \;_{2}Y_{\ell -2} { \rm e}^{-i2 \gamma} \right] \right ) \Omega_p f_\ell. \label{App8}
\end{equation}
Now we can write
\begin{eqnarray}
\;_{2}Y_{\ell m} &=& 2 \sqrt{2} N_\ell A_{\ell}^m (G^+ - G^-) {\rm e}^{im\phi} \\
\;_{-2}Y_{\ell m} &=& 2 \sqrt{2} N_\ell A_{\ell}^m (G^+ + G^-) {\rm e}^{im\phi} 
\end{eqnarray}
(see Kamionkowski \etals 1997) and so
\begin{equation}
Q_i = \sum_{\ell p} {1 \over 2}  \left ( {2 \ell + 1 \over 4 \pi} \right)^{1/2}
2 \sqrt 2 N_\ell A_{\ell}^2
(G^+ + G^-) \Omega_p f_\ell
\left [q_p \cos(2 \alpha + 2 \gamma) - u_p \sin(2 \alpha + 2 \gamma)
\right ]. \label{App10}
\end{equation}
In the limit $\ell \rightarrow \infty$, the prefactor in (\ref{App10}), tends
to 
\begin{equation}
 \left ( {2 \ell + 1 \over 4 \pi} \right )^{1/2} \sqrt{2} N_\ell A^2_\ell 
\rightarrow  \left ( {2 \ell + 1 \over 2 \pi} \right ) {1 \over \ell^4}, \label{App11}
\end{equation}
and
\begin{equation}
  G^{\pm}_{\ell m}(s) \rightarrow {1 \over 4} \ell^4 (J_0(s) \pm J_4(s)), \label{App12}
\end{equation}
and so (\ref{App10}) becomes
\begin{equation}
Q_i \approx \sum_{\ell p}   \left ( {2 \ell + 1 \over 4 \pi} \right)
J_0 \left ( (2\ell+1) \sin \beta/2 \right) \Omega_p f_\ell \left[
q_p \cos(2 \alpha + 2 \gamma) - u_p \sin (2 \alpha + 2 \gamma) \right].
 \label{App13}
\end{equation}
Summing over $\ell$,
\begin{equation}
Q_i \approx {\Omega_p \over \theta_2^2} {1 \over 2 \pi}
 \sum_p   \left[ q_p \cos(2 \alpha + 2 \gamma) - u_p \sin (2 \alpha + 2 \gamma) 
\right ] \; \rm{exp}\left ( - {\theta^2_{ip} \over 2 \theta^2_s} \right ), \label{App14}
\end{equation}
for Gaussian smoothing. Thus, if the $q_p$ and $u_p$ are uncorrelated
between pixels and are uncorrelated with each other, and for
small angular separations, it is possible to show after some algebra
that equation (\ref{App14}) gives exactly the same covariance matrix
as the scalar result of equation (\ref{App5})
\begin{equation}
\langle Q_i Q_j \rangle \approx {1 \over \theta_s^4} {1 \over 2 \pi^2}
\sum_p \sigma^2_p \Omega^2_p {\rm exp} \left ( - {\theta^2_{ip} \over 2 \theta^2_s} \right ){\rm exp} \left ( - {\theta^2_{jp} \over 2 \theta^2_s} \right). \label{App15}
\end{equation}
An analogous derivation applies for the covariance matrix $\langle U_iU_j \rangle$
(note that $\langle Q_iU_j \rangle \approx 0$). By comparing with numerical
simulations, we find that the scalar approximation (\ref{App15}) is an
excellent approximation and is perfectly adequate for the smoothing scales
adopted in this paper.
\end{appendix}

\end{document}